\documentclass[nofootinbib,aps,prd,preprint, superscriptaddress,
showpacs,amssymb]{revtex4}

\usepackage{amsmath,amsfonts}

\usepackage{graphics}
\usepackage{color}
\usepackage{dcolumn}

\allowdisplaybreaks

\newcommand{\be}{\begin{equation}}
\newcommand{\ee}{\end{equation}}
\newcommand{\bea}{\begin{eqnarray}}
\newcommand{\eea}{\end{eqnarray}}
\newcommand{\ecc}{eccentricity }
\newcommand{\bs}{\begin{subequations}}
\newcommand{\es}{\end{subequations}}
\def\no{\nonumber \\}
\def\le{\biggl (}
\def\ri{\biggr )}

\begin{document}

\title{ Phasing of gravitational waves  from  inspiralling eccentric binaries}

\author{Thibault Damour}
\affiliation{Institut des Hautes Etudes Scientifiques,
91440 Bures-sur-Yvette, France}

\author{ Achamveedu Gopakumar}

\affiliation{Theoretisch-Physikalisches Institut,
Friedrich-Schiller-Universit\"at, Max-Wien-Platz 1, 07743 Jena, Germany}
\affiliation{
Department of Physics, University of Guelph, Canada N1G 2W1}

\author{Bala R.\ Iyer}
\affiliation{Raman Research Institute, Bangalore 560 080, India}

\date{\today}

\begin{abstract}
We provide a method for analytically constructing high-accuracy
templates for the gravitational wave signals emitted by compact binaries
moving in inspiralling eccentric orbits.
By contrast to the simpler problem of modeling the gravitational
wave signals emitted by inspiralling {\it circular} orbits,
which contain only two different time scales,
namely those associated with the orbital motion and 
the radiation reaction,
the case of {\it inspiralling eccentric}
orbits involves {\it three different time scales}: orbital period,
periastron precession and radiation-reaction time scales.
By using an improved `method of variation of constants', we show how
to combine these three time scales, without making the usual
approximation of treating the radiative time scale as an adiabatic process.
We explicitly implement our method at the 2.5PN post-Newtonian
accuracy.
Our final results can be viewed as computing new `post-adiabatic'
short period contributions to the orbital phasing, or equivalently,
new short-period contributions to the gravitational wave polarizations,
$h_{+,\times}$, that should be explicitly added to the 
`post-Newtonian' expansion for 
$h_{+,\times}$, if one treats radiative effects on the orbital phasing of
the latter in the usual adiabatic approximation.
Our results should be of importance both for the
LIGO/VIRGO/GEO network of ground based interferometric 
gravitational wave detectors (especially if 
Kozai oscillations turn out to be
 significant in globular cluster triplets), and for the
future space-based interferometer LISA.

\end{abstract}


\maketitle

\section{Introduction}
\label{IntroSec}
Inspiralling black hole binaries 
are considered to be the most probable source of detectable
gravitational radiation for the  first generation
laser interferometric gravitational-wave detectors 
that are operational or nearing
the completion of their construction phase \cite{GWIFs}.
The above understanding is based on both astrophysical considerations 
\cite{GLPPS} and the availability of  highly accurate 
general relativistic theoretical waveforms required to pluck the 
weak gravitational wave signal from the  noisy interferometric  data 
\cite{DIJS}.
Inspiralling compact binaries are usually modeled as point particles
in {\it quasi-circular} orbits. For long lived compact binaries, the  quasi-circular 
approximation is quite appropriate, as the radiation reaction
decreases the orbital  \ecc to negligible values by  the  epoch the 
emitted gravitational radiation  enters the sensitive bandwidth of the interferometers.
It is easy to deduce that for an isolated binary, 
the \ecc goes down roughly  by a factor of three, when its semi-major
axis is halved \cite{P64}.

In recent times, however, scenarios involving compact 
{\it eccentric} binaries
are being suggested  as  potential gravitational wave 
sources {\em even}  for the terrestrial
gravitational-wave detectors. 
For instance, one such  proposed astrophysical scenario \cite{MH02}, 
involves hierarchical triplets (say, 123), usually
modeled to consist of an inner (say, 12) 
and an outer binary (say, 03, where 0 denotes the center of mass
of the 12 binary).
If the mutual inclination angle between the orbital
planes of the  inner and of  the outer binary is 
large enough, then the time averaged tidal force on the inner binary
may induce oscillations in its eccentricity, known in the literature 
as the Kozai  mechanism \cite{K62}.
It was shown that in globular clusters,  the inner binaries of
hierarchical triplets undergoing Kozai oscillations can merge
under gravitational radiation reaction
 \cite{MH02}.
Later, it was shown that
a good fraction 
of such systems will have 
\ecc $\sim 0.1$, when emitted gravitational radiation from these 
binaries passes through $10 \,{\rm Hz} $ \cite{W03}.

 It is also believed that  the effect of  orbital \ecc will 
have to be modeled
accurately while computing theoretical waveforms for compact binaries
relevant for the 
space-based laser interferometric gravitational-wave detector LISA \cite{LISA}. 
The above statement is supported by a recent finding that 
by observing stellar mass black hole binaries in highly eccentric orbits -
which may be common in  globular clusters - one can estimate
accurately not just the  intrinsic binary parameters like  masses and
eccentricity, but 
even the position of  the host cluster \cite {B02}.
It is also well known that 
the cosmological supermassive black hole binaries, 
embedded in surrounding stellar populations,
would be powerful gravitational wave sources for 
detectors like LISA \cite {TB76}.  
However, the question whether these binaries will be in a
 {\it quasi-circular} or an  {\it  inspiralling eccentric} orbit by the time
their gravitational waves   are detectable by 
 LISA is  not yet settled \cite{S_MOG}. 
Recently,
it was suggested that if the  Kozai mechanism
were operative,  
these supermassive black hole binaries,
in highly eccentric orbits, would merge
within the Hubble time 
\cite{BLS03}. 
Very recently, it was shown that, 
using very-long-baseline interferometer (VLBI),  the  unresolved 
core of the radio galaxy 3C 66B executes  well-defined
elliptical motions with an yearly  period, 
which was interpreted as the first direct evidence
for the detection of a supermassive black hole binary
\cite{SIMT}. The   above observation raises the interesting
possibility of  being able to  detect gravitational waves from
 supermassive black hole
binaries in  eccentric orbits using  LISA.

    Even though various versions of `ready to use'
high accuracy  search templates
for  inspiralling compact binaries with arbitrary masses 
in {\it quasi-circular} orbit exist
\cite{DIJS}, so far none is available for 
compact objects in {\it inspiralling eccentric} orbits.
Before characterizing the strategy and new results presented
in this paper, let us summarize the relevant existing literature on
the influence of 
eccentricity on the  gravitational wave polarizations, $h_+ $ and $h_{\times}$,
with and without including the gravitational radiation reaction.
After the seminal work of Peters and Mathews \cite{PM63},
it was in the context of spacecraft Doppler detection
of gravitational waves from compact binaries that
the first explicit expressions for Newtonian accurate 
gravitational wave polarization states were derived 
\cite{W87}.
Using 
the method of osculating orbital elements
and a numerical integration approach 
the effects of eccentricity and dominant radiation damping on
$h_+ $ and $h_{\times}$ was studied in \cite{LW90}. 
 First- and First-and-half-post-Newtonian 
 accurate analytic expressions for far-zone fluxes and 
gravitational wave polarizations, for compact binaries
in eccentric orbits, were computed in 
a series of papers
in \cite{GS89}.
Using Newtonian accurate orbital motion, Refs.~\cite{MBM95,MBM94}
studied the  effect on gravitational wave polarizations
of introducing by hand some secular effects either
in the longitude of the periastron \cite{MBM95}
or in the semi-major axis and eccentricity
\cite{MBM94}.
Using such waveforms, the  influence of eccentricity on the signal to noise
ratio in  gravitational wave data analysis,
was examined in \cite{MBM94,MP99, PSLR}.
These waveforms were also used to show that LISA will be sensitive to
eccentric Galactic binary neutron stars \cite{GIKPP}
and that  by measuring their periastron advance, accurate estimates
for the total mass of these binaries may be obtained \cite{S01}.
However, the widely used  gravitational wave templates,
to detect gravitational waves from  
compact binaries in {\it quasi-circular orbits},
are based on the  {\it second post-Newtonian} accurate expressions 
for $h_+ $ and $h_{\times}$, supplemented by expressions giving
 {\it adiabatic} time evolution for the orbital phase and frequency
also to the second post-Newtonian order \cite{BDIWW,BIWW96}
[However, see \cite{DIJS_N} for the (numerical)
construction of gravitational wave templates going
{\it beyond} the adiabatic approximation in the case of
{\it quasi-circular} orbits].
The second post-Newtonian order, usually referred to as the 2PN order,
gives corrections to  leading order contributions in  gravitational theory,
to  order
of $\sim (v/c)^4 \sim (G\,m\,/c^2\, r)^2$, where $m, v $ and $r$ are the
total mass, orbital velocity and  the separation of the binary.
The 2PN contribution to the gravitational waveform,
required for the construction of $h_+ $ and $h_{\times}$,
and the associated far-zone fluxes
for binaries moving in general {\it eccentric}
  orbits, in harmonic gauge,  were computed in
\cite{WW96,GI97}.
Employing the 2PN accurate generalized quasi-Keplerian 
parametrization of Damour, Sch\"afer and Wex
\cite{DS88,SW93} available in Arnowitt, Deser and Misner (ADM) coordinates, 
to represent 
relativistic motion  binaries in eccentric orbits, 
2PN corrections to the rate of decay of 
the orbital elements of the representation as well as 
the explicit expressions for $h_+ $ and $h_{\times}$ were provided 
in \cite{GI97,GI02,GI03er}.
The above mentioned expressions for $h_+$ and $h_\times$ 
represent 
gravitational radiation from an eccentric 
binary, during that stage of inspiral 
where the  gravitational radiation reaction is so small
 that the orbital parameters can be treated as 
essentially unchanging over a few orbital
periods (`adiabatic approximation'). 
The effects of eccentricity, advance of periastron and orbital inclination
on power spectrum of the dominant Newtonian part of the polarizations
were also presented in \cite{GI02}. 

The aim of this paper is to provide a method for explicitly
constructing high-accuracy waveforms emitted by compact binaries
moving in {\it inspiralling eccentric orbits}.
Compared to the  existing high-accuracy waveforms for
{\it inspiralling circular orbits} \cite{DIJS},
the inclusion of orbital eccentricity into such
templates is a non-trivial 
task as eccentricity
brings along a new physical aspect, the precession
of the periastron, and thus one must contend with precession
and radiation reaction at the same time.
In the quasi-circular case, there are two time-scales related to the
orbital period
and the radiation reaction. In the quasi-eccentric case,
one has  a
third  additional time-scale related to the precession of periastron.
The technical problem we shall tackle and solve below is that of combining,
in a consistent framework these three time scales,
without making the usual approximation of treating the
radiation-reaction time scale as an adiabatic process.
We then explicitly implement our method at the `2PN+2.5PN' accuracy,
{\it i.e.} the effect of perturbing a `2PN-accurate analytic'
description of eccentric orbits by the 2.5PN level
radiation-reaction.
It is useful to note that
the gravitational wave observations of  inspiralling compact binaries,
are  analogous to the high precision radio-wave
observations of binary pulsars. 
The latter makes use of an accurate relativistic `timing
 formula'  which requires the  solution
to the relativistic  equation of
 motion  for  a compact  binary moving  in an  elliptical orbit
while the former  demands  accurate `phasing',
i.e. an accurate mathematical modeling of the continuous time
 evolution of the gravitational waveform.
The mathematical formulation, which resulted in an accurate `timing formula',
as given in \cite{DD85,DT92}, was obtained by one of us   many years  ago
 \cite{TD83, TD_F85}.
The present work will rely on techniques from
 \cite{TD83, TD_F85} to combine the 
 mentioned above
three time scales 
 to implement post-Newtonian accurate `phasing' for compact
binaries moving in elliptical orbits.

   The plan of the paper is the following. In Sec. \ref{PhasingSec},
we outline the steps required to do the `phasing'.
The method we use to find the domain of validity of our approach is 
presented in Sec. \ref{DelineateSec}.
In Sec. \ref{MethodSec}, we formulate the procedure
required to construct time evolving $h_\times$ and $h_+$.
We apply, in Sec. \ref{2.5PNSec},  the formalism to do the 2.5PN accurate 
phasing.
Sec.\ref{PNadiabaticSec} displays the expressions required to
extend the `phasing' to higher PN orders.
Finally, in Sec. \ref{ConcludeS}, we summarize our approach, results 
and point out possible future extensions.

\section{The phasing of gravitational waveforms}
\label{PhasingSec}

 The theoretical templates for compact binaries,
required to analyze  the noisy data from the
detectors, as noted earlier, usually
consist of $h_+$ and $h_\times$, the two independent 
gravitational wave (GW) polarization states, 
expressed in terms
of the  binary's intrinsic dynamical  variables and location.
            These two basic polarization states
$h_{+} $ and $ h_{\times} $ are given by
\bs
\bea
\label{Eq1a}
h_{+} &=& \frac{1}{ 2} \biggl ( p_i\,p_j
-q_i\,q_j \biggr )\,h_{ij}^{TT}\,,\\
\label{Eq1b}
h_{\times} & = &\frac{1}{2} \biggl ( p_i\,q_j
+p_j\,q_i \biggr )\,h_{ij}^{TT} \,,
\eea
\label{Eq1}
\es
where $ h_{ij}^{TT}$, 
the transverse-traceless (TT) part of the radiation
field, is expressible in terms of a post-Newtonian expansion 
in $(v/c)$,
and where $\mathbf{p}$ and  $\mathbf{q}$
are two orthogonal unit  vectors in the plane of the sky {\it i.e.}
in the plane transverse to the radial direction linking
the source to the observer.

The TT radiation field is given, by the existing gravitational
wave generation formalisms \cite{BDIWW,BIWW96,WW96},
as a post-Newtonian expansion of the form
\be
h_{ij}^{\rm TT}=\frac{1}{c^4}\left[
h_{ij}^0+
\frac{1}{c}h_{ij}^1+
\frac{1}{c^2}h_{ij}^2+
\frac{1}{c^3}h_{ij}^3+
\frac{1}{c^4}h_{ij}^4+
\frac{1}{c^5}h_{ij}^5+
\frac{1}{c^6}h_{ij}^6+
\cdots
\right]
\label{Eq2}
\ee
where, for instance, the leading (`quadrupolar') approximation
is given (in a suitably defined `center of mass frame', see below)
in terms of the relative separation vector $\mathbf{x}$ and
relative velocity vector $\mathbf{v}$ as 
\be
\frac{1}{c^4}(h^{0}_{km}) = \frac{4\,G\, \mu }{ c^4\,R'}
{\cal P}_{ijkm}({\bf N})
\le  v_{ij} -{G\,m \over r}\,n_{ij} \ri \,,
\label{Eq3}
\ee
where $ {\cal P}_{ijkm}({\bf N}) $ is the
usual transverse traceless
projection operator projecting normal to {\bf N}, where 
${\bf N} = {\bf R'}/R'$, $R'$ being the radial distance to the binary.
The reduced mass of the binary $\mu$ is  given
by $ m_1\,m_2 /m$, where $m\equiv m_1+m_2$ 
is the total mass of the binary consisting 
of individual masses $m_1$ and $m_2$.
We also used   $v_{ij}\equiv v_iv_j,$ and  $\;n_{ij}\equiv n_in_j $,
where $v_i$ and $n_i$ are the components of 
the velocity vector ${\bf v} =d{\bf x}/dt $
and the unit relative separation vector ${\bf n} = {\bf x}/r$,  where
$ r= |{\bf x}|$.
When inserting the explicit expression of $h_{ij}^0$, and
its higher-PN analogues
 $h_{ij}^1$,
 $h_{ij}^2\cdots$ which are currently known up to $h_{ij}^4$ 
\cite{WW96,GI97,GI02},
($h_{ij}^5$ is currently being explicitly derived \cite{AI03}),
one ends up with a corresponding expression for the two
independent polarization amplitudes, Eqs.~(\ref{Eq1}), as functions of
the relative separation $r$ and the `true anomaly' $\phi$, {\it i.e.}
the polar angle of $\mathbf{x}$, and their time derivatives,
\bea
h_{+,\times}(r,\phi,\dot{r},\dot{\phi}) &=& \frac{1}{c^4}
\biggl [
h_{+,\times}^0(r,\phi,\dot{r},\dot{\phi})+
\frac{1}{c}\,h_{+,\times}^1(r,\phi,\dot{r},\dot{\phi})+
\frac{1}{c^2}\,h_{+,\times}^2(r,\phi,\dot{r},\dot{\phi})
\no
&&
+ \frac{1}{c^3}\,h_{+,\times}^3(r,\phi,\dot{r},\dot{\phi})+
\frac{1}{c^4}\,h_{+,\times}^4(r,\phi,\dot{r},\dot{\phi})+
\cdots
\biggr ]\,.
\label{Eq4}
\eea
For instance, if we follow the conventions used both in \cite{WW96}
and \cite{BIWW96}, for choosing the orthonormal triad
$\mathbf{p}$, $\mathbf{q}$, $\mathbf{N}$
(namely, $\mathbf{N}$ from the source to the observer and $\mathbf{p}$
toward the correspondingly defined `ascending' node), {\it i.e.} if we use
\be
\mathbf{x}=\mathbf{p}\,r\,\cos \phi+(\mathbf{q}\,\cos i +
\mathbf{N}\, \sin i)r\,\sin \phi,
\label{Eq5}
\ee
where $i$ denotes the inclination of the orbital plane with respect
 to the plane of the sky, the lowest-order contribution in Eq.~(\ref{Eq4})
reads
\bs
\bea
\label{Eq6a}
\frac{1}{c^4}h_{+}^0 (r,\phi,\dot{r},\dot{\phi}) 
&=&- {G\,m\,\eta \, \over c^4\,R'}\,\biggl \{
( 1 +C^2) \biggl [
\biggl ( {G\, m \over r} + r^2\,\dot \phi^2 -\dot r^2
\biggr )
\cos 2\,\phi + 2\,\dot r\, r\,\dot \phi \, \sin 2\,\phi
\biggr ]
\no
&&
+ S^2
\,\biggl [ {G\, m \over r}
-r^2\,\dot \phi^2 -\dot r^2 \biggr ]
\biggr \}\,,
\\
\label{Eq6b}
\frac{1}{c^4}h_{\times}^0 (r,\phi,\dot{r},\dot{\phi}) 
&=&- 2 {G\,m\,\eta \,C \over c^4\,R'}\,\left \{
\le {G \,m \over r} + r^2\,{\dot \phi}^2 - \dot r^2
\ri \sin 2\phi - 2 \dot r \,r\,\dot \phi \,\cos 2\phi
\right \}\,,
\eea
\label{Eq6}
\es
where $ \eta \equiv \mu /m \equiv m_1m_2/(m_1+m_2)^2, C$ and $ S$ are shorthand notations for $ \cos i$
and $ \sin i$.
The orbital phase is denoted by $\phi$, $\dot \phi = {d \phi / dt}$
and $\dot r =  {d r / dt} ={\bf n} \cdot {\bf v} $, where
$ \mathbf{v}=\mathbf{p} \,( \dot r\,\cos \phi
-r\,\dot \phi \,\sin \phi  )
+(\mathbf{q}\,\cos i +
\mathbf{N}\, \sin i)\,( \dot r\,\sin \phi + r\,\dot \phi\, \cos \phi )$.
We note that in our expressions for $h_{\times}$ and $h_+$, the
coefficients of $\cos 2\,\phi$ and $\sin 2\,\phi$
differ from those
derived in \cite{GI02} by a minus sign as in that paper
the true anomaly $\phi$
was measured from ${\bf q}$,
rather than from the line of ascending
node as done here and in \cite{BIWW96}.

Having in mind the existence of expressions such as Eq.~(\ref{Eq6})
giving the GW amplitudes $h_+$, $h_\times$ in
terms of the relative motion
$\mathbf{x}$, $\mathbf{v}$ of the binary, it is clear that
they must be supplemented by explicit expressions describing
the temporal evolution of the relative motion, {\it i.e.}
describing the explicit time dependences $r(t)$, $\phi(t)$, $\dot{r}(t)$,
and $\dot{\phi}(t)$. We  refer to as  {\it phasing}, any explicit way to
define the latter time-dependences, because it is the crucial input
needed beyond the `amplitude' expansions, given by 
Eqs.~(\ref{Eq4}), to derive
some ready to use waveforms $h_{+,\times}(t)$.

Let us note two things about the structure sketched above for
$h_{+,\times}$. First, the possibility to express the GW
polarization amplitudes in terms of only the relative motion
$\mathbf{x}$, $\mathbf{v}$, relies on the possibility to go to
a suitable center-of-mass frame. The validity of a centre-of-mass theorem
up to order $c^{-5}$	inclusive {\it i.e.} in the presence
of the leading radiation reaction was first shown in \cite{TD83}
(in harmonic coordinates).
The analogous result,
in ADM coordinates,  was obtained in \cite{GS83},
where it was shown that there existed six first integrals of the 2.5PN
equations of motion: a total momentum 
${\cal{P}}_{(5)}^i= {\cal{P}}_{(0)}^i+ c^{-2}{\cal{P}}_{(2)}^i+
c^{-4}{\cal{P}}_{(4)}^i+ c^{-5}{\cal{P}}_{(5)}^i$
and `center-of-mass' constant
${\cal{K}}_{(5)}^i\equiv {\cal {G}}_{(5)}^i- t {\cal {P}}_{(5)}^i$,
which could both be set to zero by applying a suitable
Poincar\'e transformation.
The recent obtention of (manifestly or not) Poincar\'e invariant
3PN equations of motion \cite{DJS00A,BF01} and the construction
of a corresponding complete set of 3PN conserved quantities
\cite{DJS00B,ABF02}  allows one to extend the construction of
${\cal {P}}^i$ and ${\cal{K}}^i\equiv {\cal{G}}^i - t {\cal{P}}^i$
to order $c^{-6}$ inclusive and thereby define a 3PN-accurate
center-of-mass frame \cite{BI02}.
[Note, however, that the $c^{-5}$ contribution ${\cal{G}}_{(5)}^i$
to ${\cal{G}}_{(6)}^i$  introduced by \cite{BI02}
coincides with ${\cal{G}}_{(5)}^i$ of \cite{TD83} only in the
center-of-mass frame.]
At the next PN level, the above 3PN ``conserved quantities''
will not be conserved anymore because of the 3.5PN component
of radiation reaction so that

\be
\dot{{\cal{P}}}_{(6)}^i={\cal{O}}(\frac{1}{c^7})\;,\;
\dot{{\cal{K}}}_{(6)}^i={\cal{O}}(\frac{1}{c^7})\;.
\label{Eq7}
\ee
The ${\cal{O}}(c^{-7})$ `recoil' of the center-of-mass
implied by Eqs.~(\ref{Eq7}) is expected to influence the
waveform only at the ${\cal{O}}(c^{-8})$ level.
Indeed, if we think of the binary as a GW source emitting the
`relative' signal, given by Eqs. (\ref{Eq6}), in its (instantaneous)
rest frame (namely, the above defined 3PN center of mass frame),
the time-dependent recoil of the latter rest frame
will introduce both a $\mathbf{N}\cdot \mathbf{v}_{\rm CM}/c$
Doppler shift of the phasing and a corresponding  modification
of the amplitudes $h_{+,\times}$.

Second, we should mention that the possibility to express
$h_{+,\times}$ only in terms of $r$, $\phi$ and their
time derivatives holds because we restricted ourselves to non-spinning
objects. In the presence of spin interactions, the orbital plane
is no longer fixed in space and one needs to introduce further variables,
notably a (slowly varying) `longitude of the node' $\Omega$.
Correspondingly, the polarization direction $\mathbf{p}$ 
cannot be defined anymore as the line of nodes.
We note in this respect that such a situation was  dealt with in
the problem of the timing of binary pulsars \cite{DT92} and
it might be advantageous to use the conventions used there to
define $\mathbf{p}$ and $\mathbf{q}$.
Namely, in terms of Fig. 1 of \cite{DT92}, $\mathbf{p}=\mathbf{I}_0$,
$\mathbf{q}=\mathbf{J}_0$, but note that the
binary pulsar convention uses as the third vector
$\mathbf{I}_0\times \mathbf{J}_0$, the direction
from the observer to the source. Such a convention
 being natural when one thinks of
 the actual observation of a signal somewhere on the sky, as seen by us!

Finally, one should make clear the coordinate systems that we shall
use. 
Indeed, the explicit 
functional forms for 
$h_+(r,\phi, \dot r, \dot \phi),
 h_{\times} (r,\phi, \dot r, \dot \phi) $ 
as well as 
the phasing relations 
$r(t),\phi(t), \dot r(t) $ and $ \dot \phi(t)$
depend on the coordinate system used, though the final results 
$h_+(t)$ and $h_\times (t) $ do not
[Note that $ h_{ij}^{TT}$ and therefore $h_+(t)$ and $h_\times (t) $
are {\it coordinate independent} asymptotic quantities ].
Here, one has to face a slight mismatch between the {\it harmonic}
coordinate systems in which standard GW generation formalisms
derive the amplitude expressions Eqs.~(\ref{Eq4}) above,
and the {\it ADM} coordinate systems which allow one to derive
(when neglecting radiation reaction) a rather simple
and elegant `quasi-Keplerian' form of the
general (eccentric) orbital motion \cite{DS88,SW93},
and therefore of the {\it phasing} of the GW signal.
As our work is focused on the latter {\it phasing} issue
(in presence of radiation reaction), we shall consistently work in {\it ADM}-
type coordinate systems because they will allow us to write down explicit analytical expressions for the orbital phasing $r(t)$ and $\phi(t)$.
We shall then assume that the starting amplitude expressions
are first transformed from the original harmonic-coordinates form
$h_{+,\times}( \mathbf{y_1}, \mathbf{y_2},
 \mathbf{\dot{y}_1}, \mathbf{\dot{y}_2})$ to the corresponding ADM-coordinates
ones
$h_{+,\times}( \mathbf{x_1}, \mathbf{x_2},
 \mathbf{\dot{x}_1}, \mathbf{\dot{x}_2})$.
Note that we do mean expressing $h_{+,\times}$ in terms of ADM positions and
{\it velocities}, though the use of positions and {\it momenta}
 a priori looks more natural in the ADM Hamiltonian framework.
All the formulas necessary for the transformation between the two 
systems have been worked out, at the 2PN order in \cite{DS85, DS88} 
and in \cite{DJS00B,ABF02} for the 3PN.
Note that the reduction to the center-of-mass can equivalently
be performed before or after the transformation
$(\mathbf{y_a},\mathbf{\dot{y}_a})\rightarrow
(\mathbf{x_a},\mathbf{\dot{x}_a})$
\cite{BI02}. 
Evidently, this transformation, which starts at 2PN order,
does not affect the lowest-order
expressions exhibited above, Eq.~(\ref{Eq6}).

    We have explicitly computed,
in ADM coordinates,
$h_{+,\times}^1, h_{+,\times}^2, h_{+,\times}^3 $ and $h_{+,\times}^4$,
which give PN corrections to $h_{+,\times}$ to  2PN order, 
in terms of $r,\phi,\dot{r},\dot{\phi}$, in the convention
used to obtain $h_{+,\times}^0$.
In appendix \ref{AppA}, we briefly describe the steps to get the
above corrections and sketch 
the structural forms of these corrections.

\section{Delineating the `stable' eccentric orbits and 
the `quasi-Keplerian' ones}
\label{DelineateSec}

In the case of inspiralling {\it circular} orbits,
a very important role, for the phasing of gravitational waves,
is played by the last stable (circular) orbit (LSO).
In the  zeroth approximation, circular orbits above the LSO,
and in the presence of radiation reaction, can be described
as an adiabatic sequence of circular orbits.
This approximation breaks down when the binary reaches
the LSO, at which point the orbital motion changes into a kind
of (relative) plunge. To describe the smooth transition between the
adiabatic inspiral and the plunge, one needs to use a formalism
for the orbital motion (such as the `effective one body' approach
\cite{BD99}), which goes beyond the usual, purely perturbative,
post-Newtonian approach.

In the present paper, we study inspiralling {\it eccentric}
orbits, evolving under radiation reaction, 
and we shall use a purely perturbative post-Newtonian
approach (but one which goes beyond the zeroth order,
adiabatic approximation). Such a treatment can only be valid
if we stay sufficiently above any eccentric analog of the LSO,
{\it i.e.} if we consider eccentric orbits which are `stable'
in the sense of being separated by a potential barrier from
any plunge motion.
The purpose of this section is to delineate, in the plane of the
parameters representing the two-dimensional manifold of eccentric
orbits, the locus where such orbits would cease to be bona fide
bound orbits to become plunge-type ones.
To do this, we need to use the non-perturbative effective
one body (EOB) formalism, because the transition
between bound and plunge orbits has a non-perturbative,
strong-field origin. As the rest of the paper will consider
the conservative part of the orbital motion at the second post-Newtonian
(2PN) approximation, we shall consistently use the EOB Hamiltonian
at the resummed 2PN level \cite{BD99}.
(To work at the 3PN level one should use the more complicated EOB 
formalism derived in \cite{DJS00}).

The real (resummed 2PN) EOB Hamiltonian describing the conservative part
of the orbital motion  ({\it i.e.} when neglecting radiation reaction),
has the form (we recall that $\eta\equiv \mu/m \equiv m_1m_2/(m_1+m_2)^2$)
\be
H_{\rm real}(R,P_R,P_\phi)=m c^2\sqrt{
1+2\eta\left(\widehat{H}_{\rm eff}-1\right)}\,,{\rm where},
\label{Eqn1}
\ee
\be
\widehat{H}_{\rm eff}\equiv \frac{H_{\rm eff}}{\mu c^2}=
\sqrt{A(R)\left[1+\frac{{\cal{J}}^2}{\mu^2 c^2 R^2} +\frac{P_R ^2}{\mu^2 c^2 B(R)}
\right]}\,,{\rm and},
\label{Eqn2}
\ee
\be
A(R)=1-\frac{2Gm}{c^2R}+2\eta \left(\frac{Gm}{c^2 R}\right)^3\,.
\label{Eqn3}
\ee
Here $R$ denotes the effective radial separation between the two bodies,
$P_R$ the corresponding (relative) radial momentum, and
${\cal{J}}$ the (relative) total angular momentum.
The total energy (including the rest mass)
will be denoted by ${\cal E_{\rm real}}$
(or simply ${\cal E}$ when no confusion with the `effective' 
energy can arise ).
The total energy ${\cal E_{\rm real}} = H_{\rm real}$
is related to the `effective' specific energy 
$\widehat{E}_{\rm eff} = \widehat{H}_{\rm eff} $ by
${\cal E}_{\rm real} =m\, c^2\,
\sqrt{ 1+2\eta\left(\widehat{E}_{\rm eff}-1\right)}$.
We shall not need the explicit expression of the EOB metric
component $B(R)$, but only use the fact that $B(R)>0$.
Taking into account the fact that the radial kinetic energy term
$P_R^2/(\mu^2 c^2 B(R))$ in Eq.~(\ref{Eqn2}) is positive,
the radial motion can be qualitatively understood in terms of the
angular-momentum dependent (effective) `radial potential'
\be
W_{{\cal{J}}}(R)\equiv A(R)\left[1+\frac{{\cal{J}}^2}{\mu^2 c^2 R^2}\right]\,,
\label{Eqn4}
\ee
which is a generalization (to the comparable mass case)
of the well-known radial potential for test-particle 
orbits around a Schwarzschild black hole, namely,
\be
W_{{\cal{J}}}^0(R)=(1-2 G m/c^2 R)(1+{\cal{J}}^2/(\mu^2 c^2 R^2))
\label{Eqa1}
\ee
(which is simply the $\eta\rightarrow 0$ limit of Eq.~(\ref{Eqn4})).

\begin{figure}
\resizebox{16cm}{!}{\includegraphics{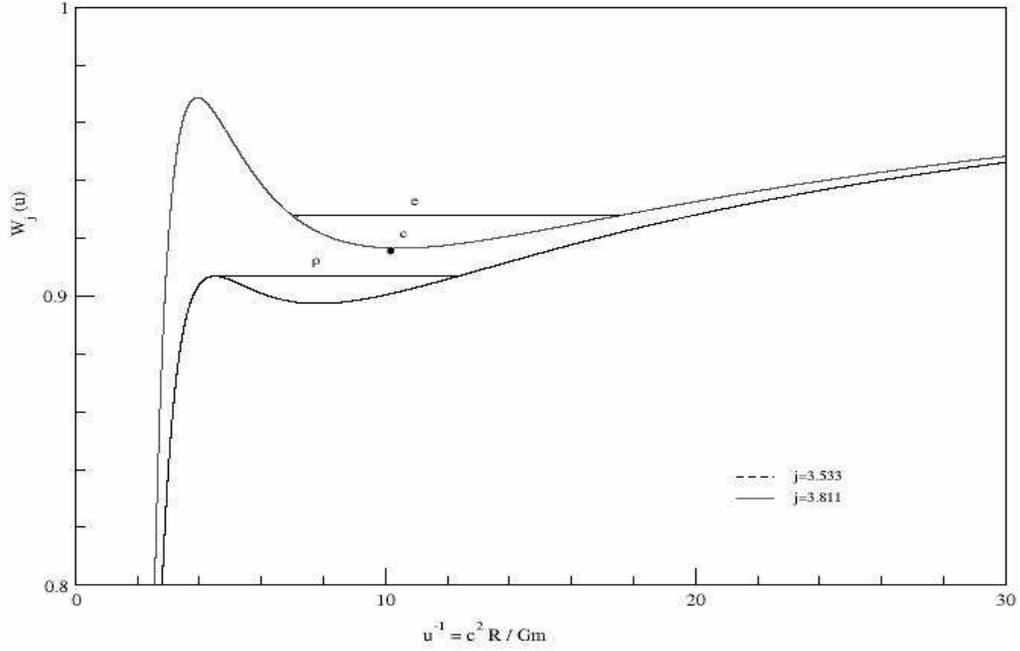}}
\caption{
The 2PN accurate effective radial potential $W_j (u)$
as a function of the
dimensionless radial variable $ \frac{1}{u}
=\frac{c^2\,R}{G\,m} $,
for various values of the dimensionless angular momentum $j$.
The point, marked c, denotes a {\it stable}
circular orbit, while the line, noted e,
stands for a  precessing elliptical orbit.
The line with label p denotes an elliptical orbit which is about to plunge.
Note that the
left end of the line p is tangent to the effective potential,
and corresponds to an {\it unstable} circular orbit.
The plots are for $\eta =0.25$.
}
\label{fig:w_u}
\end{figure}

In Fig.~\ref{fig:w_u},
we present  typical plots of 
$W_{{\cal{J}}}(R)$ on which we mark 
`energy levels' corresponding to some 
eccentric and circular orbits.
This plot makes it clear that an {\it a priori} bound motion ({\it i.e.}
with total energy ${\cal{E}}_{\rm real} < mc^2$, {\it i.e.}
$\widehat{E}_{\rm eff} < 1$) will execute some precessing but
stable motion only if the energy-level $\widehat{H}^2_{\rm eff}=
\widehat{E}^2_{\rm eff}$ stays above the radial potential
$W_{{\cal{J}}}(R)$ in a finite radial interval $\left[R_{\rm min}\,,\,R_{\rm max}\right]$.
The `centrifugal barrier' preventing this radial motion to plunge
towards smaller separations is on the left of $R_{\rm min}$.
Therefore the locus of orbits which are on the verge of plunging 
corresponds to the case where the energy-level (horizontal) line would
be tangent to the top of the potential barrier, {\it i.e.}
to
\be
\frac{dW_{{\cal{J}}}(R)}{dR}=0\,.
\label{Eqn5}
\ee
In the domain of interest, the Eq. (\ref{Eqn5}), considered
for some given ${\cal{J}}$, will have two roots, say $R_p({\cal{J}})$ and
$R_c({\cal{J}})$, with $R_p < R_c$.
The larger root $R_c({\cal{J}})$ defines the set of
{\it stable} circular orbits. Note, in this respect, that the
smaller root $R_p({\cal{J}})$, which marks the `plunge' locus also
corresponds to the locus of {\it unstable} circular orbits.
Finally, the domain in the energy-angular momentum
plane corresponding to bound (non plunge) motion is restricted
by the inequalities
\be
W_{{\cal{J}}}(R_c({\cal{J}}))<\widehat{E}^2_{\rm eff} < W_{{\cal{J}}}(R_p({\cal{J}}))\,,
\label{Eqn6}
\ee
which defines, when using the link (\ref{Eqn1}),
a corresponding double inequality involving ${\cal{J}}$ and
${\cal{E}}_{\rm real}$.
Note that, as one approaches the plunge boundary,
{\em i.e.}  for orbits close to the orbit marked `p' in 
Fig.~\ref{fig:w_u}, the character of the
orbital motion starts to deviate very much from that of a usual, perturbative,
slowly precessing, ``quasi-Keplerian" motion. Instead, it becomes
what is referred to as 
a ``zoom-whirl" motion in \cite{GK02},
{\em i.e.} a motion which alternates
between one large-excursion elliptic-Keplerian-like orbit
and several quasi-circular orbits near the periastron (which,
as noted above is close to an {\it unstable} circular orbit).
As the  formalism we shall use in this paper to analytically represent
the orbital motion assumes a ``quasi-Keplerian" representation (see below),
we shall need to stay sufficiently away from the plunge boundary to
ensure the numerical validity of such a representation. Before
coming to the issue of what one  exactly means by  ``sufficiently away",
let us finish describing the analytical estimate of the plunge boundary,
as defined by the inequalities given in Eq. (\ref{Eqn6}).

    Let us analytically estimate the two crucial roots $R_p({\cal{J}})$ 
and $R_c({\cal{J}})$
of Eq.~(\ref{Eqn5}). Using the dimensionless, scaled variables 
$u\equiv Gm/c^2 R$, $j\equiv c {\cal{J}}/(\mu G m)$, the radial potential
reads
\bs
\label{Eqn7}
\bea
\label{Eqn7a}
W(u)&=&A(u)\left(1+j^2 u^2\right)\;\\
\label{Eqn7b}
{\rm where},\;A(u)&=&1-2 u+2\eta u^3\,.
\eea
\es
The equation (\ref{Eqn5}), or better $-W'(u)/(2j^2)$, reads
\be
3u^2-u+\frac{1}{j^2}-\eta u^2(\frac{3}{j^2}+5 u^2)=0\,.
\label{Eqn8}
\ee
When $\eta\rightarrow 0$,
Eq.~(\ref{Eqn8}) becomes a quadratic equation, with
the two roots
\be
u^{\pm}_0(j)\equiv \frac{1}{6}\left[1\pm \sqrt{1-\frac{12}{j^2}}\;\;\right] \,,
\label{Eqn9}
\ee
where the plus sign corresponds to the `plunge'
boundary (larger $u$, {\it i.e.} smaller $R$),
while the minus sign corresponds to circular orbits.
An accurate (at least when $j^2 > 12$) analytical estimate
of the $\eta$-deformations
of the above two roots,
{\it i.e.} the roots of
the quartic equation, Eq. (\ref{Eqn8}), corresponding to 
$R_p$ and $R_c$, is obtained by inserting expressions (\ref{Eqn9}) into the
$\eta$-dependent terms in  Eq. (\ref{Eqn8}).
This yields
\be
u^{\pm}(j)\simeq \frac{1}{6}\left[1\pm \sqrt
{1-\frac{12}{j^2}
\left[
1-\eta(u_0^\pm)^2(3+5j^2(u_0^\pm)^2)\right]}\;\right] \,.
\label{Eqn10}
\ee
We have verified that the analytical estimate, Eq. (\ref{Eqn10}),
is a numerically
accurate estimate of the two roots $u_p(j)$, $u_c(j)$.
Inserting this result (with $u^+=u_p$, $u^-=u_c$)
into Eq. (\ref{Eqn6}) then yields an explicit (2PN-level) estimate of the
domain of `non-plunge' eccentric orbits in the 
$({\cal{E}},{\cal{J}})$ plane.
Another way to describe the plunge boundary in the 
$({\cal{E}},{\cal{J}})$ plane,
which does not need to assume that $ j^2 > 12$, 
is to give a {\it parametric}  representation of this boundary
in terms of the parameter $u$ 
in the form ${\cal{E}} = {\cal{E}}(u), {\cal{J}} = {\cal{J}}(u)$.
This is simply obtained by solving Eq. (\ref{Eqn8}) for $j$,
which gives 
$j(u) = \frac{\sqrt{1-3\,\eta\,u^2} }{ \sqrt{ u\,(1 -3\,u +5\,\eta\,u^3)}} $. One then obtains 
${\cal{E}} = {\cal{E}}(u)$ by substituting for ${\cal{J}}$ by
$\frac{j(u)\,\mu\,G\,m}{c}$ in Eqs. (\ref{Eqn1}) and (\ref{Eqn2}).

    Having discussed the location of the plunge boundary, let us now discuss
the issue of `how far from the boundary' we need to stay for allowing
us to rightfully use the analytical quasi-Keplerian representation of
\cite{DS88,SW93}
(see Eqs. (\ref{Eq12}) below).
As this representation assumes, among other
things, that the orbits are slowly precessing, we need to set an upper
limit on the rate of periastron precession. [This will then, de facto,
eliminate the possibility of whirl-zoom orbits, which contain, during
part of their orbital period, a rapidly precessing quasi-circular motion].
To see more precisely what upper limit we should set on periastron precession,
let us go back to the EOB representation
of the motion. From previous work on the 2PN-accurate EOB dynamics
\cite{BD99},
we know that the comparable mass case ($\eta\sim0.25 $)
is rather close to the test-mass one ($\eta \rightarrow 0$), yielding
geodesic motion in a Schwarzschild spacetime. 
Therefore, let us  consider
the domain of parameter space for which periastron precession
around a  Schwarzschild  black hole is well described by a slowly
precessing, quasi-Keplerian motion. For this, we  consider
the exact formula, given by Eq.~(A8) in the Appendix~A of \cite{DS88},
which gives the angle of return to the periastron for a test particle
moving 
in Schwarzschild spacetime.
It can be easily checked that,
for the elliptical orbits ( the eccentricity parameter $e_t < 1$) 
we are interested in, the term whose expansion is most slowly convergent is the
prefactor $ ( 1- \frac{12}{j^2} )^{-1/4} $ in Eq. (A8) of 
\cite{DS88}. We must therefore impose   $\frac{12}{j^2}  \ll 1 $
to have a slowly precessing, quasi-Keplerian motion.
When this inequality is
satisfied (together with $ 0 \leq e_t < 1 $), we expect that
the 2PN-accurate expressions for 
$n = \frac{2\,\pi}{T}$, $T$ being the radial orbital period,
and $e_t^2$ in terms of
${\cal{E}}$ and ${\cal{J}}$, as derived in \cite{DS88,SW93},
to be numerically accurate.
In terms of dimensionless 
non-relativistic energy per unit reduced mass 
$ E \equiv { ( {\cal E} - m\,c^2 )}/{ \mu\,c^2} $ 
and $j$,
defined earlier as $ c {\cal{J}}/(\mu G m)$,
the expressions for $n$ and $e_t^2$ read  \cite{E_def}
\bs
\bea
\label{Eq19Na}
\xi \equiv \frac{G\,m\,n}{c^3}&=&(-2\, E)^{3/2}
\biggl \{ 1 -\frac{1}{8}\left ( 15 -\eta \right ) (-2\, E)
+ \frac{ (-2\, E)^2}{128} \left ( 555 + 30\,\eta + 11\,\eta^2
\right )
\no
&&
- \frac{3}{2} \left ( 5 - 2\,\eta \right ) 
\frac{ (-2\, E)^{3/2} }{j}
\biggr \}\,,
\\
\label{Eq19Nb}
e_t^2&=&
1+2 \, E\,j^2 + 
{ E}
\biggl \{  4\,(1-\eta ) + (17 - 7\,\eta)\,  E\,j^2
\biggr \}
+
\biggl \{ 2( 2 + \eta + 5\,\eta^2 )\,  E^2
\no
&&
- (17 -11\,\eta) \frac{ E}{j^2} 
+ (112 - 47\,\eta + 16\,\eta^2 )\, E^3 j^2
\no
&&
-3(5 -2\,\eta)(1+2\, E j^2)
\frac{(-2\, E)^{3/2}}{j}
\biggr \}\,.
\eea
\label{Eq.39N}
\es
     Using the  above expressions one can approximately express $\frac{12}{j^2}$
in terms of $\xi$  and $e_t$, and define
\be
\epsilon \equiv   \frac{12}{j^2}
\sim 12 \, \frac{ \xi^{2/3} }{ (1-e_t^2)} 
\ee
We can now specify `what small means' in terms of $\epsilon$
to ensure a decent convergence of  the crucial factor
$ ( 1- \epsilon)^{-1/4}$ entering the  periastron precession expression. 
A minimal requirement would be to impose $\epsilon < \frac{1}{4}$.
Indeed, when $\epsilon = \frac{1}{4}$
the 2PN-accurate expression for the
periastron-advance parameter $k' \equiv  ( 1-\epsilon)^{-1/4} -1 $, 
(which yields the periastron advance of nearly circular orbits)
namely $k' =\frac{\epsilon}{4} +\frac{5\,\epsilon^2}{32}$
gives the exact value to an accuracy  $\sim 3\%$.
Choosing such a threshold, $\epsilon < \frac{1}{4}$, for
`staying sufficiently away from
the plunge boundary' leads to the following constraint on the
parameters $\xi$ and $e_t$:
\be
\frac{\xi}{(1-e_t^2)^{3/2} }=
\biggl ( \frac{\epsilon}{12} \biggr )^{3/2} < 3.0 \times 10^{-3}
\label{ineqA}
\ee

    Later, when we evolve orbital elements and  gravitational 
waveforms, we make sure that 
the eccentric orbits we study
lie inside this domain, defined by the above inequality.
{\em Let us emphasize that this restriction is due to 
our use of, in the next section, 
the generalized quasi-Keplerian representation,
given by Eqs. (\ref{Eq14}), (\ref{Eq15}) and (\ref{Eq16}).
We could go beyond the limit, given by Eq. (\ref{ineqA}),
by using, instead of 
the generalized quasi-Keplerian representation, the
exact Schwarzschild-like  motion (analytically expressible in terms of
rather simple quadratures) in the EOB metric.}
This will be tackled in the near future.

\section{A method of variation of constants}
\label{MethodSec}

In this section, we introduce a version of the  general Lagrange
method of variation of arbitrary constants, which was 
employed to  compute,  within   general relativity, the orbital 
evolution of the  Hulse-Taylor binary pulsar \cite{TD83,TD_F85}.
The method begins by splitting the relative  acceleration
of the compact binary ${\cal A} $ into two parts,
an integrable   leading  
part ${\cal A}_{0} $ and  a perturbation  part,  
${\cal A}'$ as
\be
\mathbf{{\cal A}}
= \mathbf{{\cal A}}_0 + \mathbf{{\cal A}'}\,.
\label{Eq8}
\ee
In this work, we will work at 2.5PN accuracy and
accordingly  choose 
$\mathbf{{\cal A}}_{0}$ to be
the acceleration at  2PN order
and $\mathbf{{\cal A}}'$ to be
 the  $c^{-5}$ (leading)
contribution to radiation reaction.
It will, however, be clear that our method is general and can be applied,
for instance, to a 3.5PN-accurate calculation where 
$\mathbf{{\cal A}}_{0}$ would be the conservative part of the 3PN
dynamics, and 
$\mathbf{{\cal A}}'$ the ${\cal{O}}(c^{-5}) + {\cal{O}}(c^{-7})$
radiation reaction.
The method
first constructs the solution to
the `unperturbed' system, defined by
\bs
\bea
\label{Eq9a}
\dot{\mathbf{x}}&=&\mathbf{v}\,,\\
\label{Eq9b}
\dot{\mathbf{v}}&=& {\cal A}_{0} (\mathbf{x},\mathbf{v}).
\eea
\label{unpert}
\es
The solution to the exact system
\bs
\bea
\label{Eq10a}
\dot{\mathbf{x}}&=&\mathbf{v}\,,\\
\label{Eq10b}
\dot{\mathbf{v}}&=& {\cal A} (\mathbf{x},\mathbf{v})\,,
\eea
\label{full}
\es
is then  obtained by 
{\it varying the constants} in the generic solutions of
the unperturbed system, given by Eqs.~(\ref{unpert}).
The method assumes (as is true for 
$\mathbf{{\cal A}}_{2PN}^{\rm conservative}$ or 
$\mathbf{{\cal A}}_{3PN}^{\rm conservative}$)
that the unperturbed system admits sufficiently many integrals of motion
to be integrable. For instance, if we work with
$\mathbf{{\cal A}}_{0}= \mathbf{{\cal A}}_{2PN}$,
we have four first integrals: 
the 2PN accurate energy and 2PN accurate
 angular momentum of the binary.
We denote these quantities, written in the 2PN accurate 
center of mass frame as $c_1$ and $c_2^i$:
\bs
\bea
\label{Eq11a}
c_1&=&
{\cal E}
(\mathbf{x_1},\mathbf{x_2},\mathbf{v_1},\mathbf{v_2})|_
{\rm 2PN~ CM}\,,\\
\label{Eq11b}
c_2^i&=&
{\cal J}_i
(\mathbf{x_1},\mathbf{x_2},\mathbf{v_1},\mathbf{v_2})|_
{\rm 2PN~ CM}\,,
\eea
\label{Eq11}
\es
with corresponding 3PN  definitions of $c_1$ and $c_2^i$, 
if we were working with
$\mathbf{{\cal A}}_{0}= \mathbf{{\cal A}}_{3PN}^{\rm conservative}$.

The vectorial structure of $c_2^i$, indicates that the  unperturbed  motion  
takes place in a plane.
The problem is restricted to a plane even in the presence of radiation
reaction \cite{TD83}.
 We may therefore introduce polar coordinates 
in the plane of the orbit $r$ and $\phi$ such that 
$\mathbf{x}= \mathbf{i}\,r \cos \phi  + \mathbf{j}\,r \sin \phi $
with, say, $\mathbf{i}=\mathbf{p}$,
$\mathbf{j}=\mathbf{q} \cos i + \mathbf{N} \sin i $ (see above).
The functional form for the solution to the   unperturbed 
equations of motion, 
following \cite{TD83,DS88}, 
may be 
expressed as 
\bs
\bea
\label{Eq12a}
r=S(l;c_1,c_2)\;\;&;&\;\;
\dot{r}=n \frac{\partial S}{\partial l}(l;c_1,c_2)\,,\\
\label{Eq12b}
\phi=\lambda +W(l;c_1,c_2)\;\;&;&\;\;
\dot{\phi}=(1+k)n + n \frac{\partial W}{\partial l}(l;c_1,c_2)\,,
\eea
\label{Eq12}
\es
where $\lambda$\footnote{We denote by $\lambda$, the variable denoted by $m$ in 
Ref. \cite{TD83}.  In 
most current literature including this one,  $m$ denotes
the total mass of the binary. Also note that, the variable $k$ here is related
to the $k$-variable in \cite{GI02} say $k_{\rm GI}$  by $k=k_{\rm GI}/c^2$}
and $l$ are two basic angles, which are  
$2\pi$ periodic
and $c_2= |c_2^i|$.
The functions $S(l)$ and $ W(l)$ and therefore 
$\frac{\partial S}{\partial l}(l)$ and 
$\frac{\partial W}{\partial l}(l)$ are periodic in $l$ with 
a period of $2\pi$. In the   above equations,
$n$  denotes the unperturbed `mean motion', given by
$n = \frac{2\pi }{P},$ $P$  being the radial
(periastron to periastron) period,
while $k=\Delta \Phi/2\pi$, $\Delta \Phi$ being
the advance of the periastron in the time interval
$P$.
The explicit 2PN accurate expressions for $P$ and $k$ in terms of
$ c_1$ and $c_2 $ are given in \cite{DS88}.
The corresponding 3PN accurate ones are given in 
\cite{DJS00B}.
The angles $l$ and $\lambda$ satisfy, still for the unperturbed system,
$\dot{l}=n$ and $\dot{\lambda}=(1+k)n$, which integrate to
\bs
\bea
\label{Eq13a}
l&=&n(t-t_0)+c_l\,,\\
\label{Eq13b}
\lambda&=&(1+k)n(t-t_0)+c_\lambda\,,
\eea
\label{Eq13}
\es
where $t_0$ is some initial instant and the constants  $c_l$ and $ c_\lambda$, 
the corresponding
values for $l$ and $\lambda$.
Finally, the unperturbed solution depends on four integration constants:
$c_1$, $c_2$, $c_l$ and $c_\lambda$.

At the 2PN order,
one can write down explicit expressions for 
the functions $S(l)$ and $W(l)$. 
Indeed, the 
generalized quasi-Keplerian representation \cite{DS88,SW93} yields:
\bs
\bea
\label{Eq14a}
S(l;c_1,c_2)&=&a_r(1-e_r \cos u )\,,\\
\label{Eq14b}
W(l;c_1,c_2)&=&(1+k)(v-l)+\frac{f_\phi}{c^4}\sin 2v +\frac{g_\phi}{c^4}\sin 3v\,,
\eea
\label{Eq14}
\es
where $v$ and $u$ are 
some 2PN accurate true and eccentric anomalies, which 
must, in Eq.~(\ref{Eq14}), be expressed as
functions of $l,\,c_1,\,$ and $c_2$, say as
$v={\cal{V}}(l;c_1,c_2)=V({\cal{U}}(l;c_1,c_2))$ and
$ u = {\cal{U}}(l;c_1,c_2) $. In the above equations, $a_r$ and $e_r$ are 
some 2PN accurate semi-major axis and radial eccentricity, while 
$f_\phi$ and $g_\phi$ are certain functions, 
given in terms of $c_1$ and $c_2$.
[To avoid introducing new notation, the eccentric anomaly is denoted by
$u$ following standard convention. It should not be confused
with $u=G m/c^2 R$ employed in Sec. \ref{DelineateSec}.
A similar comment applies to  the function $v$ below in the quasi-Keplerian
representation and the magnitude of the relative velocity $v$.]
The definitions of 2PN accurate functions $u={\cal{U}}(l;c_1,c_2)$
and $v=V(u)$ are available in \cite{DS88,SW93}.
First, the function $v\equiv V(u)$ is defined by
\be
v= V(u)\equiv 2 \arctan{ \le \le\frac{1+e_\phi}{1-e_\phi}
\ri^{1/2}\tan\frac{u}{2} \ri}\,.
\label{Eq15}
\ee
Second, the function $u={\cal{U}}(l)$  is defined by inverting the following
`Kepler equation' $l=l(u)$
\be
l=u-e_t\sin u +\frac{f_t}{c^4}\sin V(u) +\frac{g_t}{c^4}(V(u)-u) \,.
\label{Eq16}
\ee
Then the function $v={\cal{V}}(l)$ is obtained by inserting $u={\cal{U}}(l)$
in $v=V(u)$, {\it i.e.}
${\cal{V}}(l)\equiv V({\cal{U}}(l))$.
Here $e_t$ and $e_\phi$ are some time and angular eccentricity and
$f_t$ and $g_t$ are certain functions of $c_1$ and $c_2$, 
appearing at the 2PN order. 
In our computations, we use the following exact relation 
for $v-u$, which is also periodic in $u$, given by
\be
\label{Eq33a}
v -u = 2\, \tan^{-1} \biggl 
( \frac{\beta_{\phi}\,\sin u}{1-\beta_{\phi}\,\cos u}
\biggr ) \,,
\ee
where $ \beta_{\phi} = \frac{ 1 -\sqrt{1-e_\phi^2}}{e_\phi} $.
We note that the extension
of such a generalized quasi-Keplerian representation to
the 3PN order is easily possible when working in ADM type coordinates.

    Let us now turn to use
the explicit unperturbed solution,
Eqs. (\ref{Eq12}) and (\ref{Eq13}),
for the construction of the general solution of the perturbed
system, Eqs. (\ref{full}).
This is done by 
keeping exactly the same functional form for $r$, $\dot{r}$,
$\phi$ and $\dot{\phi}$,
as functions of $l$ and $\lambda$, Eqs.~(\ref{Eq12}),
{\it i.e.} by writing 
\bs
\bea
\label{Eq17a}
r=S(l;c_1,c_2)\;\;&;&\;\;
\dot{r}=n \frac{\partial S}{\partial l}(l;c_1,c_2)\,,\\
\label{Eq17b}
\phi=\lambda +W(l;c_1,c_2)\;\;&;&\;\;
\dot{\phi}=(1+k)n + n \frac{\partial W}{\partial l}(l;c_1,c_2),
\eea
\label{Eq17}
\es
but by allowing temporal variation in $c_1=c_1(t)$
and $c_2 = c_2(t)$ [with corresponding temporal variation in
$n = n(c_1, c_2)$  and $k = k(c_1, c_2) $], 
{\it and}, by modifying the unperturbed expressions, given by
Eqs. (\ref{Eq13}), for the temporal variation of the
basic angles $l$ and $\lambda$ entering Eqs. (\ref{Eq17})
into the new expressions: 
\bs
\bea
\label{Eq18a}
l &\equiv&\int_{t_0}^t n\, dt \,+\, c_l(t)\,,\\
\label{Eq18b}
\lambda&\equiv&\int_{t_0}^t (1+k)\, n\, dt \,+\, c_\lambda (t)\,,
\eea
\label{Eq18}
\es
involving two new evolving quantities $c_l(t)$, and $c_\lambda(t)$.
In other words, we seek for solutions of the exact system,
Eqs. (\ref{full}),
in the form given by Eqs. (\ref{Eq17}) and (\ref{Eq18}) with {\it four}
`varying constants'
$c_1(t)$, $c_2(t)$, $c_l(t)$ and  $c_\lambda(t)$.
The four variables $\{c_1,c_2,c_l,c_\lambda\}$ replace the original
four dynamical variables $r$, $\dot{r}$, $\phi$ and $\dot{\phi}$.
It can be verified 
that the alternate set $ \{ c_1$,
$c_2$, $c_l, c_\lambda \}$
satisfies,
like the original
phase-space variables, first order evolution equations
\cite{TD83,TD_F85}.
These evolution equations have a rather 
simple functional  form, namely,
\be
\frac{dc_\alpha}{dt}=F_\alpha(l;c_\beta )\;;\, \alpha,\beta =1,2,l,\lambda\,,
\label{Eq19}
\ee
where the right hand side is linear in the perturbing acceleration,
${\cal A}'$. 
Note the presence of the sole angle $l$ (apart from the implicit
 time dependence
of $c_\beta$) on the RHS of Eqs. (\ref{Eq19}).
The explicit expressions for these evolution equations were  derived 
in \cite{TD_F85}, which in our notation read
\bs
\bea
\label{Eq20a}
\frac{d c_1}{dt} &=& \frac{ \partial c_1(\mathbf{x},\mathbf{v}) }
{\partial v^i}\,
{ \cal  A'}^{i} \,,\\
\label{Eq20b}
\frac{d c_2}{dt} &=& \frac{\partial c_2(\mathbf{x},\mathbf{v})}
{\partial v^j }\,
{\cal A'}^{j} \,,\\
\label{Eq20c}
\frac{d c_l}{dt} &=& - \le  \frac{\partial S}{ \partial l} \ri ^{-1} 
\le  \frac{\partial S}{ \partial c_1}\, \frac{d c_1}{dt}
+ \frac{\partial S}{ \partial c_2}\, \frac{d c_2}{dt} 
\ri \,, \\
\label{Eq20d}
{ d c_\lambda \over dt} &=&
-\frac {\partial W } {\partial l} {d c_l \over dt} 
-\frac {\partial W } {\partial c_1} {d c_1 \over dt} 
- \frac {\partial W } {\partial c_2} {d c_2 \over dt} \,.
\eea
\label{Eq20}
\es
The evolution equations for 
$c_1$ and $c_2$ clearly  arise from
 the fact that $c_1$ and $c_2$ 
were defined as some first integrals in phase-space,
say Eqs.~(\ref{Eq11}).
As shown in \cite{TD_F85}, there is an alternative
expression for $\frac{d c_l}{dt}$, which reads
\be
\frac{d c_l}{dt} = 
\le \frac{\partial Q}{ \partial l} \ri ^{-1} 
\le  {\cal A'} \cdot  {\bf n} - 
\frac{\partial Q}{ \partial c_1}\, \frac{d c_1}{dt}
- \frac{\partial Q}{ \partial c_2}\, \frac{d c_2}{dt} \ri \,,
\label{Eq21}
\ee
where $Q^2(l, c_1, c_2) = \dot r^2 (S(l,c_1,c_2), c_1, c_2 ) $ and 
$ \frac{\partial Q}{ \partial l} $ is defined by 
\be
\frac{\partial Q}{ \partial l} = \frac{P}{4\,\pi} 
\frac{ \partial Q^2}{\partial r}\,.
\label{Eq22}
\ee
Both expressions, Eqs. (\ref{Eq20c}) and (\ref{Eq21}),
involve formal delicate limits of the $0/0$ form, for some
(different) values of $l$. Taken together, they prove that
these limits are well-defined and yield for $dc_l/dt$,
an everywhere regular function of $l$.
Anyway, the algebraic manipulation of the explicit forms of both
Eqs. (\ref{Eq20c}) and  (\ref{Eq21})
 lead to well-defined expressions 
[For example,
in the case of Eq. (\ref{Eq20c}),
the problematic $\sin u$ factor in $\partial S/\partial l$,
see Eq.~(\ref{Eq32a}) below, nicely simplifies with a $\sin u$ factor
present in the term within parenthesis on the RHS of
Eq. (\ref{Eq20c})].

The definition of $l$ given by $ l= \int_{t_0}^t n (c_a (t)) 
\, dt \,+\, c_l(t) $,   is equivalent to the differential form, 
$\frac{dl}{dt}=n + \frac{d c_l}{dt} = n + F_l (l,c_a); a =1\,,\,2 $, which allow us to define 
a set of differential equations for $c_\alpha$ as
functions of $l$ similar
to Eqs.~(\ref{Eq19}) for $c_\alpha$ as functions of $t$.
The exact form of the differential equations for $c_\alpha(l)$ reads
\be
\frac{dc_\alpha}{dl}=\frac{F_\alpha(l;c_a)}
{n(c_a)+ F_l(l;c_a)},
\label{Eq23}
\ee
where $c_a$, $a=1,2,$ stands for $c_1$ and $c_2$.
Neglecting terms quadratic in $F_\alpha$, {\it i.e.} quadratic
 in the perturbation $\mathbf{{\cal{A}}}'$
(e.g. neglecting ${\cal{O}}(c^{-10})$ terms in our application),
we can simplify the system above to 
\be
\frac{dc_\alpha}{dl} \simeq  \frac{1}{n(c_a)}\,F_\alpha(l;c_a)
\equiv  G_\alpha (l; c_a)
\;;\, \alpha =1,2,l,\lambda\,; a= 1\,,\,2\,.
\label{Eq24}
\ee
From here onward, we will neglect these $ {\cal{O}}(c^{-10})$ terms 
in the evolution equations for $c_\alpha (l) $, {\it i.e.} work with the
simplified system, namely Eq. (\ref{Eq24}).
At this stage, it is crucial to note, not only that the RHS of Eq.~(\ref{Eq24})
is a function of $c_1$, $c_2$ and the sole angle $l$ (and not of $\lambda$),
but that it is a {\it periodic} function of $l$.
This periodicity,
together with the slow [$G_\alpha \propto F_\alpha \propto \mathbf{{\cal{A}}}'=
{\cal{O}}(c^{-5})$] evolution of the $c_\alpha$'s, implies
that the evolution of $c_\alpha (l)$
 contains both a `slow' (radiation-reaction time-scale)
 secular drift and `fast' (orbital time-scale) 
periodic oscillations.
For the purpose of phasing, to model the combination of
 slow drift and the fast oscillations
present in  $c_\alpha$,
we introduce a two-scale decomposition for $ c_\alpha (l) $ in the 
following manner
\be
c_\alpha (l)=\bar{c}_\alpha (l)+\tilde{c}_\alpha (l)\,,
\label{Eq25}
\ee
where the first term $\bar{c}_\alpha (l)$ represents a slow drift
(which can ultimately lead to {\it large} changes in the `constants' $c_\alpha$)
 and $\tilde{c}_\alpha (l)$ represents fast oscillations
(which will stay always {\it small}, {\it i.e.} of order
${\cal{O}}(G_\alpha)={\cal{O}}(c^{-5})$).
This is proved by first decomposing the periodic functions
$G_\alpha(l)$ (considered for fixed values of the other arguments
$c_a$
) into its {\it average} part and its {\it oscillatory}
part:
\bs
\bea
\label{Eq26a}
\bar{G}_\alpha (c_a)&\equiv&\frac{1}{2\pi}\int _0^{2\pi}\, dl\, G(l,
c_a)\,,\\
\label{Eq26b}
\tilde{G}_\alpha (l;c_a)&\equiv&G_\alpha (l;c_a)-\bar{G}_\alpha (c_a).
\eea
\label{Eq26}
\es
Note that, {\it by definition}, the oscillatory part
$\tilde{G}_\alpha(l)$ is a periodic function with {\it zero average}
over $l$.
Then assuming that $\tilde{c}_\alpha$ in Eq.~(\ref{Eq25}) is always
small ($\tilde{c}_\alpha={\cal{O}}(G_\alpha)={\cal{O}}(c^{-5})$),
one can expand the RHS of the exact evolution system, 
given by Eqs. (\ref{Eq24}), as
\bea
\label{Eq27}
\frac{d\bar{c}_\alpha}{dl}+
\frac{d\tilde{c}_\alpha}{dl}&=&
G_\alpha(l;\bar{c}_a+\tilde{c}_a)=
G_\alpha(l;\bar{c}_a)+
{\cal{O}}(G_\alpha^2)\,,\no
&=&\bar{G}_\alpha(l;\bar{c}_a)+
\tilde{G}_\alpha(l;\bar{c}_a)
+{\cal{O}}(G_\alpha^2)\,.
\eea
We can then solve, modulo ${\cal{O}}(G_\alpha ^2)$, the evolution equation
(\ref{Eq27}) by defining $\bar{c}_\alpha(l)$ as a solution of
the `averaged system'
\be
\label{bareq}
\frac{d\bar{c}_\alpha}{dl}= \bar{G}_\alpha(\bar{c}_a)\,,
\ee
and by defining $\tilde{c}_\alpha(l)$ as a solution of the
`oscillatory part' of the system
\be
\label{tileq}
\frac{d\tilde{c}_\alpha}{dl}= \tilde{G}_\alpha(l, \bar{c}_a)
\,.
\ee
During one orbital period ($0\leq l\leq 2\pi$) 
the quantities
$\bar{c}_a$
on the RHS of Eq.~(\ref{tileq}) change only by
${\cal{O}}(G_\alpha)$.
Therefore, by neglecting again terms of order ${\cal{O}}(G_\alpha ^2)\sim
{\cal{O}}(c^{-10})$ in the evolution of $\tilde{c}_\alpha$,
we can further define $\tilde{c}_\alpha(l)$ as the {\it unique zero-average}
solution of Eq.~(\ref{tileq}), considered for
fixed values of $\bar{c}_a$, {\it i.e.}

\be
\tilde{c}_\alpha(l) =
\int dl\, \tilde{G}_\alpha
(l;\bar{c}_a)|_{\bar{c}_a=\bar{c}_a(l)}
= \int \frac{dl}{n}\, \tilde{F}_\alpha
(l;\bar{c}_a)\,.
\label{tilsol}
\ee
The indefinite integral in Eq. (\ref{tilsol}) is defined as the unique
zero-average periodic primitive of the zero-average (periodic)
function $\tilde{G}_\alpha(l)$.
During that integration, the arguments $\bar{c}_a$
are kept fixed, and, after the integration, they are replaced 
by the slowly drifting solution of the averaged system,
given by Eqs. (\ref{bareq}).
Note that 
Eq. (\ref{tileq})
 yields $\tilde{c}_\alpha={\cal{O}}(G_\alpha)$,
which was assumed above, thereby verifying the consistency of the
(approximate) two-scale integration method used here.

  We are now in a position to apply the  above described 
method of variation of arbitrary constants, 
which gave us the evolution equations for
$\bar c_\alpha$ and $\tilde c_\alpha$, to GW phasing.
We  use 2PN accurate expressions for
the dynamical variables $r$, $\dot r$, $\phi$ and $\dot \phi$
entering the expressions for $h_{\times} $ and $h_+$, given by
Eqs. (\ref{Eq6}).
To do the phasing, we will
solve the evolution equations for
 $ \{ c_1, c_2, c_l, c_\lambda \} $, given by Eqs.~(\ref{Eq24}), on 
the 2PN accurate orbital dynamics, given in Eqs.~(\ref{Eq12}).
This leads to an evolution system,
given by Eqs. (\ref{bareq}) and (\ref{tileq}),
in which the RHS contains terms of order
${\cal{O}}(c^{-5})\times\left[1+{\cal{O}}(c^{-2})
+{\cal{O}}(c^{-4})\right]=
{\cal{O}}(c^{-5})+
{\cal{O}}(c^{-7})+
{\cal{O}}(c^{-9})$.
In the next section, as a first step, we will restrict our attention to
the leading order contributions to $\bar G_\alpha$ and $\tilde G_\alpha$,
which define the evolution of $\{ \bar c_\alpha, \tilde c_\alpha \}$ under
gravitational radiation reaction to ${\cal{O}}(c^{-5} )$ order.
We then impose these variations, via Eqs. (\ref{Eq17}) and (\ref{Eq18}),
on to $h_{\times} $ and $h_+$, given by
Eqs. (\ref{Eq6}).
This will allow us to obtain 
gravitational wave polarizations, which are Newtonian accurate in
their amplitudes and 2.5PN accurate in orbital dynamics.
We name the above procedure 2.5PN accurate phasing of gravitational waves.
Since $\tilde G_\alpha$'s create only  {\em periodic
2.5PN corrections} to the dynamics, in this paper, we will not 
explore its higher PN corrections.
However, in a later section, we will present the consequences
of considering PN corrections to $\bar G_\alpha$ by computing
${\cal{O}}(c^{-9}) $ contributions to relevant
$\frac{d \bar c_\alpha}{dt}$.
This is required as $\bar G_\alpha$ directly contribute to the
highly important adiabatic evolution of $h_{\times} $ and $h_+$.

Up to now we have assumed, for concreteness,
 that the two constants $c_1$ and $c_2$
were the energy and the angular momentum, respectively. 
However, any functions of these conserved quantities can do as well. In view of our use of the generalized quasi-Keplerian representation to describe
the orbital dynamics, it is convenient to follow \cite{GI02}
and to use as  $c_1$  the  mean motion $n$, and as $c_2$ the
time-eccentricity $e_t$. This can be done 
by employing the  2PN accurate expressions for $n$ and $e_t$
in terms of ${\cal E}$ and ${\cal J}$ (or rather E and j), derived
in \cite{DS88,SW93}.
Firstly, this will require us to express 2PN accurate orbital dynamics
in terms of $l, n$ and $e_t$.
Secondly, 
using $n$ and $e_t$, instead of 
${\cal{E}}$ and  ${\cal{J}}$, as $c_1$ and $c_2$,
we need to derive the evolution equations for $ \frac{d n }{dt} $, 
$ \frac{d e_t }{dt} $,$ \frac{d c_l }{dt} $ and $ \frac{d c_\lambda }{dt} $
in terms of $l, n $ and $e_t$. 
This will follow straightforwardly from Eqs.~(\ref{Eq20}).
Using these expressions, the  evolution equations,
namely Eqs. (\ref{bareq}) and (\ref{tileq}), for
$ \{ \bar n, \bar e_t, \bar c_l, \bar c_\lambda, 
\tilde n, \tilde e_t, 
\tilde  c_l,  \tilde c_\lambda  \} $ 
will be obtained
in terms of $l, n$ and $e_t$.
  
    As mentioned earlier, we  restrict in this paper
 the conservative dynamics to 
the 2PN order.  Below, we  present the  2PN accurate orbital dynamics,
given by Eqs.~(\ref{Eq17}), explicitly in terms of $(l, n, e_t)$.
This straightforward computation  employs  explicit expressions 
for the orbital elements of generalized quasi-Keplerian representation,
in terms of
$E$ and $j$ 
available in \cite{DS88,SW93}.
The relations we need are:
\bs
\bea
a_r(n,e_t)&=& \biggl ( \frac{G\,m}{n^2} \biggr )^{1/3}
\biggl \{ 1 -\frac{\xi^{2/3}}{3} (9 -\eta)
+ \frac{\xi^{4/3}}{72}\,\biggl [ (72 + 75\,\eta + 8\,\eta^2 
\no
&&
-\frac{1}{(1-e_t^2)^{1/2}} ( 360 - 144\,\eta ) 
-\frac{1}{(1-e_t^2)} ( 306 - 198\,\eta ) 
\biggr ]
\biggr \}\,,
\label{Eq31a}
\\
\label{Eq31b}
e_r(n,e_t)&=&
e_{{t}} \biggl \{ 1
+{ \xi^{2/3} \over 2}\,\left( 8 -3\,\eta \right)
+ { \xi^{4/3} \over 24\, (1 -e_t^2)^{3/2}}\,
\biggl [
\biggl (
 -(288
 -242\,\eta
 +21\,{\eta}^{2}
 ){e_{{t}}}^{2}
 \no
 &&
 +390
 -308\,\eta
 +21\,{\eta}^{2}
 \biggr )\,\sqrt { 1 -e_t^2}
 + \left ( 180 -72\,\eta \right )\,( 1 -e_t^2)
 \biggr ]
 \biggr \}\,,
\\
\label{Eq31c}
e_{\phi}(n,e_t) &=& e_{{t}} \biggl \{
1+{\xi}^{2/3}\left (
4-\eta\right )+
{ \xi^{4/3} \over 96\, (1 -e_t^2)^{3/2}}\,
\biggl [
\biggl (
-(1152
-656\,\eta
+41\,{\eta}^{2}
 ){e_{{t}}}^{2}
 \no
 &&
 +1968
 -1088\,\eta
 -4\,{\eta}^{2}
 \biggr )\sqrt {1 -e_t^2}
 + \left ( 720 -288\,\eta \right )\,( 1 -e_t^2)
 \biggr ]
 \biggr \} \,,
\\
k(n,e_t)  &=& { 3\, \xi^{2/3} \over ( 1 -e_t^2)}+
{\xi^{4/3} \over 4\, (1 -e_t^2)^2 }
\biggl \{ ( 51 -26\,\eta)\,e_t^2 + (78 -28\,\eta) \biggr \}\,,
\label{Eq31d}
\\
\label{Eq31e}
f_t(n,e_t)&=&
-\frac{\xi^{4/3} c^4}{8\, \sqrt {1 -e_t^2} }(4 + \eta)\,\eta\, e_t\,,
\\
\label{Eq31f}
g_t(n,e_t)&=&
\frac{3\,\xi^{4/3} c^4}{2\, \sqrt {1 -e_t^2} }(5 -2\, \eta)\,,
\\
\label{Eq31g}
f_\phi(n,e_t)&=&
\frac{\xi^{4/3} c^4}{8\, (1 -e_t^2)^2 }(1 -3\,\eta)\,\eta\, e_t^2\,,
\\
\label{Eq31h}
g_\phi(n,e_t)&=&
-\frac{3\,\xi^{4/3} c^4}{32\, (1 -e_t^2)^2 }\,\eta^2\, e_t^3\,,
\eea
\label{Eq31}
\es
where $\xi \equiv \frac{G m n}{c^3}$.
We note that the generalized quasi-Keplerian orbital elements,
given in terms of $E$ and $j$ in \cite{DS88,SW93,E_def}, can easily be 
expressed in $n$ and $e_t$  using following 2PN accurate relations for
$-2\,E$ and $-2\,E\,j^2$, which read
\bs
\label{inp2}
\bea
{-2\,E}&=&{\xi}^{2/3}\biggl \{
1+{ {\xi}^{2/3} \over 12}
\biggl [ 15-\eta
\biggr ]
+{{\xi}^{4/3} \over 24\,}
\biggl[
\biggl ( 15 -15\,\eta -\eta^2
\biggr )
\no
&&
+
{1 \over \sqrt{1-e_t^2}}\,
\biggl ( 120 -48\,\eta \biggr )
\biggr ]
\biggr \} \,,
\\
-2\,E\,j^2 &=& 
(1 -e_t^2)\,\biggl \{
1+{{\xi}^{2/3} \over 4\,(1 -e_t^2) }\biggl [ -\left (
17-7\,\eta\right ){{ e_t}}^{2}
+9 +\eta  \biggr ]
\no
&&
+ { {\xi}^{4/3} \over 24\,(1 -e_t^2)^2}
\biggl [
-\left (360-144\,\eta\right ){{e_t}}^{2}
\sqrt {1 -e_t^2}+\left (225-277\,\eta+29\,{\eta}^{2}\right ){{ e_t
}}^{4}
\no
&&
-
\left (210-190\,\eta+30\,{\eta}^{2}\right ){{ e_t}}^{2}+
189-45\,\eta+{\eta}^{2}\biggr ]  \biggr \}\,.
\eea
\es
These two relations  easily follow from inverting the 
2PN accurate relations for
the orbital period $P = \frac{2\,\pi}{n}$ and $e_t^2$ in terms of
$E$ and $j$ presented in Eqs. (\ref{Eq.39N}) above [ See \cite{DS88,SW93} ].

 In addition, to  compute expressions for $\dot r$ and $\dot \phi$, we use
the following relations
\bs
\bea
\label{Eq32a}
\frac{\partial S}{\partial l}&=& a_r\, e_r\, \sin u\,
 \frac{\partial u}{\partial l}\,,\\
\label{Eq32b}
 \frac{\partial W}{\partial l}&=& \left[\left(1+k+
 \frac{2f_\phi}{c^4}\cos 2v +\frac{3g_\phi}{c^4}\cos 3v\right)
 \frac{\partial v}{\partial u}\frac{\partial u}{\partial l}
 -\left(1+k\right)\right]\,,\\
\label{Eq32c}
 \frac{\partial u}{\partial l}&=&\le 1-e_t \cos u -\frac{g_t}{c^4}+
 \frac{1}{c^4}\left(f_t\cos v +g_t\right)\frac{\partial v}{\partial u}\ri^{-1}
\,,\\
\label{Eq32d}
 \frac{\partial v}{\partial u}&=& \frac{(1-e_\phi^2 )^{1/2}}{1-e_\phi \cos u}\,.
 \eea
\label{Eq32}
\es
 The radial motion, defined by $r (l, n, e_t) $ and $\dot r (l, n, e_t) $,
reads (both in the compact form 
 and in 2PN-expanded form)
\bs
\bea
\label{Eq34a}
r &=& S(l,n, e_t)
= a_r(n,e_t)(1-e_r(n,e_t) \cos u)
\no
&&
=({G\,m \over n^2})^{1/3}\, (1 -e_t\,\cos u)\,\biggl \{ 1
- { {\xi}^{2/3} \over 6\,(1 -e_t\,\cos u)} \biggl [ \left (6
-7\,\eta\right ){e_t}\,\cos u+18-2\,\eta\biggr ]
\no
&&
+
{{\xi}^{4/3} \over 72\,\sqrt { (1 -e_t^2)^3}\,(1 -e_t\,\cos u)}
\biggl [
\biggl (
-(72-231
\,\eta+35\,{\eta}^{2} )
(1 -{{ e_t}}^{2})\,e_t\,\cos u
\no
&&
-(
72
+75\,\eta
+8\,{\eta}^{2}
 ){{ e_t}}^{2}
 -234
 +273\,\eta
 +8\,{\eta}^{2}
 \biggr )\,\sqrt {1 -e_t^2}
 \no
 &&
 -36\, (1 -e_t^2)\, (5 -2\,\eta)\,( 2 +e_t\,\cos u)
 \biggr ]
\biggr \} \,,
\\
\label{Eq34c}
{\dot r} &=& n\frac{\partial S}{\partial l} (l,n, e_t)
= { { (G\,m\,n)}^{1 \over 3}\, \over
{(1 -e_t\,\cos u)}}\,
{{ e_t}\,\sin u }
\biggl \{
1 +{ {\xi}^{2/3} \over 6} \left (6-7\,
\eta \right )
\no
&&
+{\xi^{4/3} \over 72}\,{ 1 \over ( 1 -e_t\,\cos u)^3}
\biggl [ \biggl (
-\left (72-
231\,\eta+35\,{\eta}^{2}\right )\,(e_t\,\cos u)^3
\no
&&
+\left (216
-693\,\eta
+105\,{\eta}^{2}
\right )
\,(e_t\,\cos u)^{2}
+\left (324 +
513\,\eta-96\,{\eta}^{2}
\right )\,{e_t}\,\cos u
\no
&&
- \left ( 36+9\,{\eta}\right )\,\eta\,{{e_t}}^{2} -468
-15\,\eta+35\,{\eta}^{2} \biggr )
\no
&&
+{36 \over \sqrt {1 -e_t^2} }\,
\biggl ( \left( 1-{ e_t}\,\cos u\right )^{2}\left (4-{ e_t}\,\cos u
\right )\left (5-2\,\eta \right )\biggr )
\biggr ]
\biggr \}\,.
\eea
\label{Eq34}
\es
 In the above equation,
the eccentric anomaly $u = {\cal U } (l, n, e_t)$ is given  by inverting
the 2PN accurate Kepler equation, Eq. (\ref{Eq16}), connecting $l$ and $u$,
{\it i.e.} in explicit form
\bea
{l} &=& u-{e_t}\,\sin u-
{{\xi}^{4/3} \over 8\,\sqrt {1-e_t^2}}
\,{1 \over (1 -e_t\,\cos u) }\,
\biggl \{
e_t\,\sin u\,\sqrt {1-e_t^2}\,\eta\,\left (4 +\eta\right )
\no
&&
+12\, ( 5 -2\,\eta)\,( u - v) \left ( 1 -e_t\,\cos u \right )
\biggr \}\,.
\label{Eq35}
\eea

      The angular motion, described in terms of 
$\phi$ and $\dot \phi$,
 is given by
\bs
\label{Eq36}
\bea
\label{Eq36a}
\phi(\lambda,l) &=& \lambda  + W(l) \\
\label{Eq36b}
W(l)& =& 
v -u +{e_t}\,\sin u
+ { 3\,{\xi}^{2/3} \over (1 -e_t^2)} \biggl \{  v -u +
e_t\,\sin u \biggr \} 
\no
&&
+ { \xi^{4/3} \over 32\, (1 -e_t^2)^{5/2}}
\,{ 1 \over ( 1 -e_t\,\cos u)^3 }\,
\biggl \{ 
\biggl [  
4\,\sqrt {1 -e_t^2}\,\left (1-e_t\,\cos u \right )^{2}
\biggl (\biggl \{  -(102
\no
&&
-52\,\eta ){{ e_t}}^{2}
-156
+56\,\eta
\biggr \}\,e_t\,\cos u
+\eta\,\left (4+\eta\right ){{ e_t}}^{4}+
\left (
102-60\,\eta -2\,{\eta}^{2}
\right ){{ e_t}}^{2}
\no
&&
+ 156 -52\,\eta+{
\eta}^{2}
\biggr )
+\left (1-e_t^{2}\right )\,
\biggl ( \left ( (3\, e_t^{2}
+12)\,\eta -8
\right ){(e_t\,\cos u)}^{2}
\no
&&
+\left (\left (8-6\,\eta\right )
{ e_t}^{2} +8
-24\,\eta\right )(e_t\,\cos u)
-12\,\eta\,{ e_t}^{4}
-\left (8-
27\,\eta\right ){ e_t}^{2}
\biggr )
\no
&&
\,\eta
\biggr ]\,e_t\, \sin u
+ ( 1 -e_t\,\cos u)^3\, 
\biggl [ 
48\,{(1-e_t^2)}^{2}\left (5-2\,\eta\right )
-8\,
\biggl ((51-26\,\eta ){ e_t}^{2}
\no
&&
+78-28\,\eta\biggr )
\biggr ]\, u  
+ 8\,\left (1-{ e_t}\,\cos u\right )^{3}\,\biggl [ 
\biggl (
\left (51-26\,\eta\right ){{e_t}}^{2}
\no
&&
+78-28\,\eta
\biggr )
\,\sqrt {1-e_t^2}
-(30 - 12\,\eta)\,(1 -2\,e_t^2 +e_t^4)
\biggr ]\,v 
\biggl \}
\,,
\eea
\es
\bea
{\dot \phi }&=&
{n\,\sqrt {1-e_t^2} \over \left (1-{e_t}\,\cos u\right )^{2}}
\biggl \{
1+ {\xi^{2/3} \over (1 -e_t^2)\,(1 -e_t\,\cos u) }
\biggl [
 \left (1-\eta
 \right ){e_t}\,\cos u
 \no
 &&
 -\left (4-\eta\right ){{e_t}}^{2}+3
 \biggr ]
 +
{{\xi}^{4/3} \over 12}\,{1 \over (1 -e_t\,\cos u)^3}
\biggl  [ { 1 \over (1 -e_t^2)^{3/2}}
\biggl \{
 18\,
 \left (1-{e_t}\,\cos u\right )^{2}\,
 \no
 &&
 \left ({e_t}\,\cos u-2\,{{e_t}}^{2} +1 \right )
 \left (5-2\,\eta
 \right )
 \biggr \}
 +{1 \over  {(1-e_t^2)}^2}
 \biggl \{ \biggl (
 -(9 -
 19\,\eta-14\,{\eta}^{2}
 ){{e_t}}^{2}
 \no
 &&
 -36
 +2\,\eta
 -8\,{\eta}^{2}
  \biggr ) \,(e_t\,\cos u)^3
+ \biggl (
-(48
-14\, \eta+17\,{\eta}^{2}){{e_t}}^{4}
\no
&&
+\left ( 69-79\,\eta+4\,
{\eta}^{2}\right ){{e_t}}^{2}
+114 +2\,\eta -5\,{\eta}^{2}
 \biggr )\,(e_t\,\cos u)^2
 \no
 &&
 +\biggl (
 -(6-32\,\eta-{\eta}^{2} ){{e_t}}^{4}
 +\left (93
 -19\,\eta
 +16\,{\eta}^{2}
 \right )e_t^2
 \no
 &&
-222
+50\,\eta+{\eta}^
{2}  \biggr )\,( e_t\,\cos u)
-6\,\eta\, (1 -2\,\eta )\,e_t^6
\no
&&
+\left (54
-28\,\eta-20\,{\eta}^{2}\right ){{
e_t}}^{4}-\left (
153
-61\,\eta
-2\,{\eta}^{2}
\right )
{{e_t}}^2
+144-48\,\eta
\biggr \} \biggr ]  \biggr \} \,.
\label{Eq37}
\eea
The explicit form of $\dot{\phi}$ above  follows from Eq.~(\ref{Eq17b}).

In addition to the above explicit expressions, we also need to evaluate
the RHS of 
Eqs.~(\ref{Eq20a})-(\ref{Eq20b}) and, in particular, the 2PN
accurate partial derivatives of $n$ and $e_t$ with respect to the relative
velocity ${\bf v}$.
To get these, one could combine Eqs. (\ref{Eq.39N}) with the expressions for
$E$ and $j$ in terms of relative position and velocity, 
rather than in terms of
position and  momenta as is usual in the  ADM formalism.
To the desired 2PN order, one may either start from the ordinary Lagrangian
$L ({\bf x}, {\bf v} )$ in ADM coordinates ( See  \cite{DS85} for the explicit
construction of this Lagrangian ) or (simply)
by inverting the basic Hamiltonian equation $ {\bf v} = d H/d{\bf p} $,
to get ${\bf v}$ in terms of ${\bf p} $.
However, in the next section, we  require  expressions for
$E$ and $j$ only to the well known Newtonian order.

\section{2.5PN accurate phasing}
\label{2.5PNSec}

Let us recall that our method is general and can be applied, in principle,
to any PN accuracy. For instance, we could study the effect of the
 ${\cal {O}}(c^{-5})+ {\cal {O}}(c^{-7})$
radiation reaction on the 3PN conservative motion. However, in this
work, we limit ourselves, for simplicity, to considering
the effect of the ${\cal{O}}(c^{-5})$ radiation reaction on
the 2PN motion. Accordingly, we shall, each time it is possible, 
truncate away all effects that would correspond to the ${\cal{O}}(c^{-7})$
level or beyond.
As we shall see, this approximation is probably sufficient for
{\it oscillatory} effects (in the sense of the decomposition, given 
in Eq. (\ref{Eq25}) ), which are the primary focus of this paper.
We shall discuss below, 
how our method also justifies
the usual way of deriving the {\it secular} effects linked to the
radiation reaction, and we shall obtain more accurate expressions for them.

  This section begins by providing inputs necessary for 
computing evolution equations  for  the set $ \{ {\bar c_\alpha, 
\tilde c_\alpha} \}$, to the 2.5PN order, where the index $
\alpha = n, e_t, c_l,$ and $ c_\lambda $.
 As just said, we require 
${\cal A'}$
to 2.5PN order for this purpose. 
The 2.5PN expressions for 
${\cal A'}$ 
will have to be in the  ADM gauge,
as our conservative 2PN dynamics is given in  the same gauge.
The expression for relative reactive acceleration, to 2.5PN order,
in the  ADM gauge, available in \cite{GS85}, reads
\be
{\cal A'}^{i} =
-\frac{8\,G^2\,m^2\, \eta}{15\,c^5\,r^3}
\biggl \{ -3 \biggl [ 12\, v^2 -15\,\dot r^2 + 2\, \frac{G\,m}{r}
\biggr ] \dot r\, n^i
+ \biggl [ 11\, v^2 -24\,\dot r^2 + \frac{G\,m}{r}
\biggr ]\,v^i \biggr \}\,,
\label{Eq40}
\ee
where $v^2= \dot r^2 + r^2\,\dot \phi^2$.
At this point a nice technical simplification occurs.
Though our formalism consistently combines a 2PN-accurate,
precessing motion with 2.5PN radiation reaction, the RHS's of
Eqs.~(\ref{Eq24}) are technically given by the product of
${\cal{O}}(c^{-5})$
 reaction terms by orbital expressions
given as explicit PN expansions 
${\cal{O}}(c^{0}) +{\cal{O}}(c^{-2}) +{\cal{O}}(c^{-4})$.
Therefore, if we decide, in a first approach, to neglect
${\cal{O}}(c^{-7})$
contributions to the phasing, we can simplify the RHS's of Eqs.~(\ref{Eq24})
by keeping only the leading terms in the orbital expressions. This formally
means that it is enough to use {\it Newtonian-like} approximations
for all orbital expressions appearing in Eqs. (\ref{Eq24}).
 For instance, we can simply use
$r\simeq (G M/n^2)^{1/3}(1-e_t \cos u)$,
$n\simeq (-2 {E_{\rm DS} })^{3/2}/ G\,m \simeq 
\left[(2 G m/r -v^2)\right]^{3/2}/ G\,m$,
etc. in Eqs.~(\ref{Eq24}).
Note, however, that this does not at all mean that we are approximating
the orbital motion as being a non-precessing Newtonian ellipse.
In all expressions where they are needed, we must retain the full PN expansion.
For instance, in the contribution $\lambda$, given by Eq.~(\ref{Eq51b}),
(see below)
to $\phi=\lambda +W(l)$, we must keep the 2PN accuracy for the precession rate $n(1+k)$, and augment it by the effect of the time variation
of $n(t)$ and $k(n(t),e_t(t))$, as  discussed there.

Finally, the leading  evolution equations for $ \{ \frac{dn}{dl}, 
\frac{d e_t}{dl},\frac{d\,c_l}{dl},\frac{d\,c_\lambda }{dl} \}$
in terms of $u(l,n,e_t), n,$ and  $e_t$, follow as

\bs
\bea
\label{Eq41a}
 \frac{d\,n}{dl} &=&
    -\frac{ 8\,\xi^{5/3}\,n\,\eta}{5}
     \biggl \{
     \frac{4}{\,{\chi}^{3}}- \frac{21}{\,{\chi}^{4}}
     -{\frac {4 -30\,{{e_t}}^{2}}{{\chi }^{5}}}
     +{\frac {54(1 -\,{{e_t}}^{2})}{{\chi}^{6}}}
     \no
      &&
      -{\frac {45(1-
      {{e_t}}^{2})^2}{{\chi}^{7}}}
       \biggr \}
        +{\cal{O}}(\frac{1}{c^{7}})
	 \,,\\
 \label{Eq41b}
 \frac{d\,e_t}{dl} &=&
 \frac{8\,\xi^{5/3}\,\eta\,(1-e_t^2)}{15\,e_t}
 \biggl \{
 \frac{17}{\,{\chi}^{3}}
 -\frac{46}{\,{\chi}^{4}}
 +{ \frac { 20 + 6\,{{e_t}}^{2}}{{\chi}^{5}}}
 +{\frac {54(1- {{e_t}}^{2})}{{\chi}^{6}}}
 \no
 &&
 -{\frac { 45(1 - {{e_t}}^{2})^2}
 {{\chi}^{7}}}
 \biggr \}
 +{\cal{O}}(\frac{1}{c^{7}})
\,,\\
\label{Eq41c}
\frac{d\,c_l}{dl} &=&
\frac{8\,\xi^{5/3}\, \eta\,
\sin u  }{15\,e_t}
\biggl \{
8\,{\frac {{{e_t}}^{2}}{{\chi}^{3}}}+{\frac {17-43\,{{e_t}}^
{2}}{{\chi}^{4}}}
-\frac {29  + 22\,e_t^2  -51\,{{e_t}}^{4}}
{{\chi}^{5}}
\no
&&
-{\frac {9(1 - {{e_t}}^{2})^2}{{
\chi}^{6}}}
+{\frac {45(1-{{e_t}}^{2})^3
}{{\chi}^{7}}}
\biggr \}
+{\cal{O}}(\frac{1}{c^{7}})
\,,
\\
\label{Eq41d}
\frac{d\,c_\lambda}{dl} &=&
\frac { 8\,\xi^{5/3}\,\eta\,\sin u
}{ 15\,{e_t}}
\biggl \{
\biggl [  -\frac{17}{\,{\chi}^{4}}
+{\frac { 29 + 9\,{{e_t}}^{2}}{{\chi}^{5}}}+
{\frac {9(1-{{e_t}}^{2})}{{\chi}^{6}}}
 -{\frac {45(1 - {{e_t}}^{2})^2}
 {{\chi}^{7}}}
 \biggr ]  \sqrt {(1-e_t^2)}
 \no
 &&
 +8\,{\frac
 {{{e_t}}^{2}}{{\chi}^{3}}}+{\frac {17-43\,{{e_t}}^{2}}{{\chi}^{4}}}
 - {\frac {29 + 22\,{e_t}^2-51\,{{e_t}}^{4}}{{\chi}^{5}}}
 - {\frac {9(1-{{e_t}}^{2})^2}{{\chi}^{6}} }
 \no
 &&
 +{\frac {45(1-{{e_t}}^{2})^3
 }{{\chi}^{7}}}
 \biggr \}
+{\cal{O}}(\frac{1}{c^{7}})\,,
\eea
\label{Eq41}
\es
where
$\chi \equiv (1-e_t\,\cos u),$
$\xi \equiv \frac{G\,m\,n}{c^3}$ and $ u = u(l,n,e_t)$.
We are in a position to explore the secular and periodic variations of
$\{ c_\alpha \}$ to $ {\cal O} (c^{-5})$, which will be done in the next two
subsections.

\subsection{Secular variations}
\label{SecularSec}
   Using the  above set of equations with Eqs.~(\ref{bareq}) and (\ref{tileq}),
we obtain the differential equations for
 $\{ \bar c_\alpha, \tilde c_\alpha\}$, where the index  $\alpha\, =\, n, e_t, 
 c_l, c_\lambda  $.
Let us first consider the secular variations 
of $c_\alpha$ given by Eqs. (\ref{bareq}).
One remark is that, 
after using
an $l$-variable formulation to
separate the secular variations from the oscillatory ones, we can, at
the end, re-express the secular result, Eqs. (\ref{bareq}), in terms of the original
time variable $t$. This leads to
\be
\frac{d\bar{c}_\alpha}{dt}= 
\bar{F}_\alpha (\bar{c}_a)
\label{Eq42}
\ee
where $\bar{F}_\alpha$ is the $l$-average of the RHS of the $t$-variation
of the $c_\alpha$'s, see Eq.~(\ref{Eq19}):
$\bar{F}_\alpha(\bar{c}_a)=(2\pi)^{-1}\int _0^{2\pi}
 dl\,F_\alpha(l;\bar{c}_a)$.
Among the secular variations, let us first discuss the secular variation
of $c_l$ and $c_\lambda$.
The `source term' for the secular variations of $c_l$ and $c_\lambda$ is
the $l$-average of $F_l$ or $F_\lambda$, {\it i.e.} modulo a (secular)
factor $n(\bar{c}_a)$, the $l$-averages of $G_l=F_l/n$,
or $G_\lambda=F_\lambda/n$, {\it i.e.}
the $l$-average of the RHS's of Eqs.~(\ref{Eq41c}) and (\ref{Eq41d}).
A look at the RHS's show that, being of the form
$\sin u\, f(\cos u)$, they are odd under $u\rightarrow -u$, so
that their average over $dl\simeq (1- e_t \cos u) du$ exactly {\it vanishes}:
$\bar{G}_l=0=\bar{G}_\lambda$.
In fact,
this remarkable finding follows 
from the time-odd character
of the perturbing force $\mathbf{{\cal{A}}}'$, and therefore
would also hold if we considered the radiation reaction to the accuracy 
${\cal{O}}(c^{-5}) + {\cal{O}}(c^{-7})+ {\cal{O}}(c^{-8})
 + {\cal{O}}(c^{-9}) $.
[ We stop at $c^{-9}$ order because of the conceptual 
subtleties arising in the meaning of radiation reaction
at $c^{-10}$ order, which is the first order where nonlinear
effects linked to the leading 
${\cal{O}}(c^{-5})$ radiation reaction enter].
Note that the ${\cal{O}}(c^{-8})$ contribution to radiation
reaction comes from the {\it tail } contributions, which in its exact form
is given by an integral over the past \cite{BD88}.
This correction is {\it time-reversal asymmetric} without being simply
{\it time-reversal antisymmetric}.
However, when approximating that integral as a function of the
instantaneous state, 
it becomes a time-reversal antisymmetric
function of $\mathbf{x}$ and $\mathbf{v}$ \cite{BS93}.
Indeed, $c_1$ and $c_2$ being even under time-reversal, the partial
derivatives $\partial c_1/\partial v^i$,
$\partial c_2/\partial v^i$,
in Eqs.~(\ref{Eq20}) are time-odd.
When ${\cal {A}}'$ is time-odd, Eqs.~(\ref{Eq20}) then imply
that $dc_1/dt$ and $dc_2/dt$ are time-even.
Then in Eq.~(\ref{Eq20c}), $\partial S/\partial l$ is time-odd
(because $r$ is even, but $l$ is odd), $\partial S/\partial c_a$ is even
and $dc_a/dt$ is time-even, so that $dc_l/dt$ ends up being time-odd.
The same conclusion is found to hold for $dc_\lambda/dt$, thereby
ensuring the absence of secular variations for both $c_l$ and
$c_\lambda$:
\bs
\bea
\label{Eq43a}
 \frac{d\, {\bar c_l} }{dt} = 0&;&
\bar{c}_l(t)=\bar{c}_l(0)
\,,\\
\label{Eq43b}
 \frac{d\,{\bar c_\lambda}}{dt} = 0 &;& 
\bar{c}_\lambda(t)=\bar{c}_\lambda(0)
\,.
\eea
\label{Eq43}
\es
Turning now to the secular variations of $\bar{n}$ and $\bar{e}_t$,
Eq.~(\ref{Eq42}), we note that they reduce to the
usual `adiabatic' estimate of the secular variation
of constants, namely
\be
\frac{d\bar{c}_a}{dt}=\langle \frac{\partial c_a(\mathbf{x},\mathbf{v})}
{\partial v^i} {\cal{A'}}^i\rangle_l
\label{Eq44}
\ee
where $\langle\rangle_l$ denotes an average over an (instantaneous) orbital period.
If we were using as $c_1$ and $c_2$ the system's dimensionless energy 
$E$ and angular momentum $j$, Eq.~(\ref{Eq44})
is the usual way of estimating the secular change of $E$
and $j$ under the influence of a perturbing acceleration
 $\mathbf{{\cal{A}}}'$.
Applying Eq.~(\ref{Eq44}) to the case where $c_1=n$ and
$c_2=e_t$ is easily seen to lead simply to a coupled
differential system for $\bar{n}(t)$, $\bar{e}_t(t)$,
which is strictly equivalent (under the map
 $\bar{n}=\bar{n}( E,j)$,
 $\bar{e}_t=\bar{e}_t( E,j))$
to the just mentioned secular evolution system for
$ E$ and $ j$.
The $l$-average of the RHS of Eqs.~(\ref{Eq41a}) and (\ref{Eq41b})
in the  leading ${\cal {O}}(c^{-5})$ approximation,
as mentioned earlier, only lead to the leading terms in
the secular evolution of $\bar{n}$ and $\bar{e}_t$.
Since the RHS's of Eqs.~(\ref{Eq41a}) and (\ref{Eq41b})
are expressed in terms of $u$, it is convenient to do the 
orbital average by expressing it as an integral over
 using $u$, using $dl\simeq (1-e_t \cos u) du$.
The resulting definite integrals may be easily computed, using   
\cite{WW27}, which give
\be
\frac{1}{2\,\pi} \int_{0}^{2\,\pi}
\frac{du}{ (1-e_t\,\cos u)^{N+1}}
= \frac{1}{(1-e_t^2)^{(N+1)/2}}\, P_{N} 
\biggl ( \frac{1}{ \sqrt {1-e_t^2}} \biggr )\,,
\label{Eq.55n}
\ee
where $P_N$ is the Legendre polynomial.
Using Eq. (\ref{Eq.55n}) in
Eqs.~(\ref{Eq41a}) and (\ref{Eq41b}), we obtain
leading ${\cal {O}}(c^{-5})$ corrections to 
$\frac{d \bar n}{dl} $ and $\frac{d \bar e_t}{dl} $.
From these expressions, using $dl = \bar n \, dt $,
we obtain  $\frac{d \bar n}{dt} $ and $\frac{d \bar e_t}{dt} $,
which read
\bs
\label{Eq.57n}
\bea
\frac{d \bar n}{dt} &=&
\frac { (G\,m)^{5/3}\,n^{11/3}\,\eta}
{5\,c^5\, (1 -e_t^2)^{7/2}}
\biggl \{  96
+292\,{{e_t}}^{2}
+37\,{{e_t}}^{4} \biggr \}
+ {\cal{O}}(\frac{1}{c^7})
\,,\\
\frac{d \bar e_t}{dt} &=&
-\frac { (G\,m\,n)^{5/3}\,n\,\eta \, e_t }
{ 15\,c^5\, (1 -e_t^2)^{5/2}}
\biggl \{ 304 +  121\,{{e_t}}^{2} \biggr \}
+ {\cal{O}}(\frac{1}{c^7})
\,.
\eea
\es
These results are equivalent to the old results of Peters \cite{P64}
based on balance argument between far-zone fluxes and
local radiation damping.
In the next section, we will present 
2PN accurate  expressions, providing  
corrections to ${\cal {O}}(c^{-10})$
for $\frac{d \bar n}{dt}$ and $\frac{d \bar e_t}{dt}$.

            Let us finally note that formally one can  analytically
 solve the coupled evolution system by
successive approximations, reducing it to simple quadratures. For instance, at the leading order where one keeps only the ${\cal{O}}(c^{-5})$
contributions, one can first eliminate $t$ by dividing
$d\bar{n}/dt$ by $d\bar{e}_t/dt$, thereby obtaining an equation of the form
$d\ln \bar{n}=f_0(\bar{e}_t)d\bar{e}_t$.
Integration of this equation yields 
\be
\bar n ( \bar e_t) = 
n_i\, 
\frac { e_i^{18/19}\,
(304 + 121\,e_i^2)^{1305/2299} }{(1-e_i^2)^{3/2}} 
\, \frac{ ( 1-e_t^2)^{3/2}}{ e_t^{18/19}\, (304 + 121\,e_t^2)^{1305/2299}}\,,
\label{Eq48}
\ee
where $e_i$ is the value of $e_t$ when $n=n_i$
a result 
first obtained by Peters in \cite{P64}.

Inserting Eq.~(\ref{Eq48}) into the leading evolution equation for $\bar{e}_t$
then leads to an evolution of the form $d\bar{e}_t/dt=g_0(\bar{e}_t)$
which can be done by quadrature:
$t=\int d\bar{e}_t\,g_0^{-1}(\bar{e}_t)+ {\rm constant}$.
We can then insert back the leading result, Eq.(\ref{Eq48}), into
the leading correction terms of the evolution equation for
 $d\ln \bar{n}/d\bar{e}_t$
to get again a decoupled equation of the form $d\ln \bar{n}=f_2(\bar{e}_t)
d\bar{e}_t$ which can be integrated.
Continuing in this way would give the function $\bar{n}(\bar{e}_t)$
in the form of a expansion, which would lead to an explicit
decoupled equation for the temporal evolution of $\bar{e}_t$:
$d\bar{e}_t/dt=g(\bar{e}_t)=g_0+c^{-2}g_2$, which can again be solved 
by quadrature. This procedure may easily be extended to 
${\cal O}(c^{-9})$ order.
At the leading 2.5PN order, we have checked that the temporal evolution for
$(\bar n, \bar e_t) $, obtained by solving coupled differential
equations, Eq. (\ref{Eq.57n}), 
is in excellent agreement with those
given by the above mentioned procedure.

\subsection{Periodic variations}
\label{PeriodicSec}
Let us turn to the differential equations, which give at
${\cal O}(c^{-5})$ order, 
orbital period oscillations 
to our dynamical variables.
They read
\bs
\label{Eq49}
\bea
\label{Eq49a}
 \frac{d\,\tilde n}{dl} &=&
 -\frac{ 8\,\xi^{5/3}\,n\,\eta}{5}
\biggl \{
\frac{4}{\,{\chi}^{3}}- \frac{21}{\,{\chi}^{4}}
-{\frac {4 -30\,{{e_t}}^{2}}{{\chi }^{5}}}
+{\frac {54(1 -{{e_t}}^{2})}{{\chi}^{6}}}
\no
 &&
-{\frac {45(1- {{e_t}}^{2})^2}{{\chi}^{7}}}
 \biggr \}
- \frac { \xi^{5/3}\,n\,\eta}
{5\, (1 -e_t^2)^{7/2}}
\biggl \{  96
+292\,{{e_t}}^{2}
+37\,{{e_t}}^{4} \biggr \}
 \,,\\
 \label{Eq49b}
\frac{d\,\tilde e_t}{dl} &=&
\frac{8\,\xi^{5/3}\,\eta\,(1-e_t^2)}{15\,e_t}
\biggl \{
\frac{17}{\,{\chi}^{3}}
-\frac{46}{\,{\chi}^{4}}
+{ \frac { 20 + 6\,{{e_t}}^{2}}{{\chi}^{5}}}
+{\frac {54(1-{{e_t}}^{2})}{{\chi}^{6}}}
\no
&&
-{\frac { 45(1 - {{e_t}}^{2})^2
}{{\chi}^{7}}}
\biggr \}
+ \frac { \xi^{5/3}\,\eta \, e_t }
{ 15\, (1 -e_t^2)^{5/2}}
\biggl \{  304 + 121\,{{e_t}}^{2}  \biggr \}
\,,\\
\label{Eq49c}
\frac{d\, \tilde c_l}{dl} &=&
\frac{8\,\xi^{5/3}\, \eta\,
\sin u  }{15\,e_t}
\biggl \{
8\,{\frac {{{e_t}}^{2}}{{\chi}^{3}}}+{\frac {17-43\,{{e_t}}^
{2}}{{\chi}^{4}}}
-\frac {29  + 22\,e_t^2  -51\,{{e_t}}^{4}}
{{\chi}^{5}}
\no
&&
-{\frac {9(1-{{e_t}}^{2})^2}{{
\chi}^{6}}}+{\frac {45(1-{{e_t}}^{2})^3
}{{\chi}^{7}}}
\biggr \}
\,,
\\
\label{Eq49d}
\frac{d\, \tilde c_\lambda}{dl} &=&
\frac { 8\,\xi^{5/3}\,\eta\,\sin u
}{ 15\,{e_t}}
\biggl \{
\biggl [  -\frac{17}{\,{\chi}^{4}}
+{\frac { 29 + 9\,{{e_t}}^{2} }{{\chi}^{5}}}+
 {\frac {9(1-{{e_t}}^{2})}{{\chi}^{6}}}
 -{\frac { 45(1 - {{e_t}}^{2})^2}{{\chi}^{7}}}
  \biggr ]  \sqrt {(1-e_t^2)}
  \no
  &&
  +8\,{\frac
 {{{e_t}}^{2}}{{\chi}^{3}}}+{\frac {17-43\,{{e_t}}^{2}}{{\chi}^{4}}}
 -{\frac {29  +  22\,{{e_t}}^{2}
 -51\,{{e_t}}^{4} }{{\chi}^
     {5}}}
 -{\frac {9(1-{{e_t}}^{2})^2}{{\chi}^{6}} }
\no
&&
+{\frac {45(1-{{e_t}}^{2})^3
}{{\chi}^{7}}}
\biggr \}\,,
\eea
\es
where $n$ and $e_t$, on the right hand side of these equations,
again stand for $\bar n$ and $\bar e_t$.
Here the RHS's of Eqs.~(\ref{Eq49}) are {\it zero-average}
oscillatory functions of $l$.
[The RHS's for Eqs. (\ref{Eq49c}) and (\ref{Eq49d}) are in fact 
identical to the ones of Eqs.~(\ref{Eq41c}) and (\ref{Eq41d}),
in view of our previous result $\bar{G}_l=\bar{G}_\lambda=0$,
except for the fact that they are expressions in terms of
$\bar{n}$ and $\bar{e}_t$, instead of $n$ and $e_t$.]

           One can analytically integrate Eqs.~(\ref{Eq49}) to get
$\tilde{n}$,
$\tilde{e}_t$,
$\tilde{c}_l$,
$\tilde{c}_\lambda$
as {\it zero average} oscillatory functions of $l$.
We find, when 
expressed in terms of $u$

\bs
\label{Eq50}
\bea
\label{Eq50a}
\tilde{n}&=&
\frac { \xi^{5/3}\, n\, e_t\,\eta\,
\sin u }{15\, (1-e_t^2)^3}
\biggl \{
{\frac { 602 + 673\,{{e_t}}^{2} }{\chi}}
+{\frac {314 -203\,{{e_t}}^{2} -111\,{{e_t}}^{ 4}
}{{\chi}^{2}}}
\no
&&
+{\frac { 122 -196\,{{e_t}}^{2} +26\,{{e_t}}^{4}
+ 48\,{{e_t}}^{6 }
}{{\chi}^{3}}}
+{\frac {162(1-{{e_t}}^{2})^3
}{{\chi}^{4}}}
\no
&&
+{\frac { 216(1 -{{e_t}}^{2})^4 
}{{\chi}^{5}}}
\biggr \}
\no
&&
+\frac { \xi^{5/3} \, n\,
\eta }{ 5\, (1-e_t^2)^{7/2}}
\,\left(
96 +292\,{{e_t}}^{2} + 37\,{{ e_t}}^{4}
\right)
\biggl \{
2\,\tan^{-1} \left( {\frac {\beta_t\,\sin u }{1-\beta_t\,\cos u
}} \right) +{e_t}\,\sin  u \biggr \}
\,,
\\
\label{Eq50b}
\tilde{e}_t&=&
-\frac { \xi^{5/3}\,\eta\,\sin u
 } { 45\, (1-e_t^2)^2 }
 \biggl \{
 {\frac {134 + 1069\,{{e_t}}^{2}+72\,{{e_t}}^{4}}{\chi}}+{
 \frac {
 134+157\,{{e_t}}^{2} -291\,{{e_t}}^{4}  }{{\chi}^{2}}}
 \no
 &&
 -{\frac {
 70  -380\,{{ e_t}}^{2} +550\,{{e_t}}^{4}
 -240\,{{e_t}}^{6}
 }{{\chi}^{3}}}
 +{\frac { 162(1 -{{e_t}}^{2})^3 
 }{{\chi}^{4}}}
 \no
 &&
 +{\frac { 216(1 - {{ e_t}}^{2})^4
 }{{\chi}^{5}}}
 \biggr \}
 \no
 &&
-\frac { \xi^{5/3}\,e_t\,\eta }
 {15\, (1-e_t^2)^{5/2}}
 \left(304 +  121\,{{e_t}}^{2} \right)
 \biggl \{
 2\,\tan^{-1} \left( {\frac {\beta_t\,\sin u }{1-\beta_t\,\cos u
 }} \right) +{e_t}\,\sin u
  \biggr \}
  \,,\\
\label{Eq50c}
\tilde{c}_l&=&
\frac { 2\,\xi^{5/3}\,\eta}{45\, e_t^2}
\biggl \{
-96\,{\frac {{{e_t}}^{2}}{\chi}}
-{\frac {102-258\,{{e_t}}^{2}}{{\chi}^{2}}}
+{\frac { 116 + 88\,{{e_t}}^{2}-204\,{{e_t}}^{4}
}{{\chi}^{3}}}
\no
&&
+{\frac { 27(1 - {{e_t}}^{2})^2 
}{{\chi}^{4}}}
-{\frac {
108(1 -{{e_t}}^{2})^3
}{{\chi}^{5}}}
\no
&&
-\frac {1}{2\, (1-e_t^2)^{3/2}}
\biggl [
-134 +281\,{{e_t}}^{2} +
315\,{{e_t}}^{4}
\biggr ]
\biggr \}
\,,\\
\label{Eq50d}
\tilde{c}_\lambda&=&
\frac {2\, \xi^{5/3}\,\eta}{45\,e_t^2}
\biggl \{
\biggl [
102\, \frac{1}{{\chi}^{2}}
-{\frac { 116 +36\,{{e_t}}^{2}}{{\chi}^{3}}}
-{\frac {27(1 - {{e_t}}^{2})}{{\chi}^{4}}}
\no
&&
+{\frac { 108(1 
-{{e_t}}^{2})^2
}{{\chi}^{5}}}
\biggr ]\,\sqrt {1-e_t^2}
-96\,{\frac {{{e_t}}^{2}}{\chi}}
-{\frac {102 - 258\,{{e_t}}^{2}}{{\chi}^{2}}}
\no
&&
+{\frac { 116 + 88\,{{e_t}}^{2}-204\,{{e_t}}^{4}
}{{\chi}^{3}}}
+{\frac { 27(1 -{{e_t}}^{2})^2 
}{{ \chi}^{4}}}
\no
&&
-{\frac {
108(1 - {{e_t}}^{2})^3
}{{\chi}^{5}}}
-\frac{1}{2\,(1-e_t^2)^{9/2}}
\biggl [
\biggl ( 
-134
+ 281\,{{e_t}}^{2}
\no
&&
+ 315\,e_t^4
 \biggr )\, ( 1-e_t^2)^3 + \biggl (
134 +175\,{{e_t}}^{2} + 45\,{{e_t}}^{4}
\biggr )
\, (1-e_t^2)^{5/2}
\biggr \}
\,,
\eea
\es
where $\beta_t = \frac{ 1 -\sqrt{1-e_t^2}}{e_t} $.
Finally, let us consider the way the previous results feed in to
the basic angles $l$ and $\lambda$ entering our
perturbed solution Eq.~(\ref{full}).
From the definitions for 
$l(t)$ and $\lambda(t)$, as given in Eqs. (\ref{Eq18}),
we see that we can also split these angles in  `secular'
pieces, say $\bar{l}$, $\bar{\lambda}$ and `oscillatory' ones
$\tilde{l}$, $\tilde{\lambda}$, 
as
\bs
\label{Eq51}
\bea
\label{Eq51a}
l(t)&=&\bar{l}(t)+\tilde{l}(l;\bar{c}_a(t))\\
\label{Eq51b}
\lambda(t)&=&\bar{\lambda}(t)+\tilde{\lambda}(l;\bar{c}_a(t))\,,\,{\rm where}\\
\label{Eq51c}
  \bar{l}(t)&\equiv&\int_{t_0}^t\bar{n}(t)dt +\bar{c}_l(t) \,,{\rm and}\\
\label{Eq51d}
\bar{\lambda}(t)&\equiv&\int_{t_0}^t(1+ \bar{k}(t))\bar{n}(t)dt
+\bar{c}_\lambda(t).
\eea
\es
We note, based on earlier results, that 
$\bar{c}_l(t)=\bar{c}_l(t_0)$ and 
$\bar{c}_\lambda(t)=\bar{c}_\lambda(t_0)$ are constants.
The `oscillatory' contributions to $l$ and $\lambda$ are
given by
\bs
\bea
\label{Eq52a}
\tilde{l}(l;\bar{c}_a)&=&\int dl \frac{\tilde{n}(l)}{n}+\tilde{c}_l(l)\\
\label{Eq52b}
\tilde{\lambda}(l;\bar{c}_a)&=&\int dl\left[\frac{\tilde{n}}{n}+
\bar{k}\frac{\tilde{n}}{n}+\tilde{k}\right]+\tilde{c}_\lambda (l)
\eea
\es
Here $\tilde{k}\equiv (\partial k/\partial n)\tilde{n}+
(\partial k/\partial e_t) \tilde{e}_t$
denotes the oscillatory piece in $k$ and $\int dl \tilde{f}(l)$
denotes the unique {\it zero average} primitive of the
zero-average periodic function of $l$, $\tilde{f}(l)$.

To complete our study of the ${\cal {O}}(c^{-5})$
oscillatory contributions to the phasing, we see from Eq.~(\ref{Eq52a})
that we need to integrate $\tilde{n}/n$ and add it to the
previous result for $\tilde{c}_l(l)$.
[Note that they appear together
in the phasing formula.]
We find
\bea
\tilde{l}(l;\bar{c}_a) &=&
\frac { \xi^{5/3}\,\eta}{15\,(1-e_t^2)^3}
\biggl \{
\left( 602 +
673\,{{e_t}}^{2} \right) \chi
+ \left(
314 
-203\,{{e_t}}^{2}
-111\,e_t^4
\right) \ln \chi
\no
&&
- \left ( 602 +  673\,{{e_t}}^{2} \right )
- {\frac { 122 - 196\,{{e_t}}^{2} +26\,{{e_t}}^{4}
+48\,{{ e_t}}^{6}
}{\chi}}
\no
&&
-{\frac {
81(1 - {{e_t}}^{2})^3
}{{\chi}^{2}}}
- { \frac { 72(1 - {{e_t}}^{2})^4
}{{\chi}^{3}}}
\biggr \}
\no
&& +
\frac {\xi^{5/3}\,\eta\,}{5\,(1-e_t^2)^{7/2}}
\left( 96 +292\,{{e_t}}^{2} +
37\,{{ e_t}}^{4}
\right)
\biggl \{
\no
&&
\int \! \left(2\,\tan^{-1} \left( {\frac {\beta_t\,\sin \left( u \right)
}{1-\beta_t\,\cos \left( u \right) }} \right) +{e_t}\,\sin \left( u
\right)  \right)\,\chi\, du
\biggr \} + \tilde c_l (l)
\,,
\label{Eq53}
\eea
where $\tilde c_l (l)$ is given by
Eq. (\ref{Eq50c}) and $\beta_t = \frac{ 1 -\sqrt{1-e_t^2}}{e_t} $.
The explicit expression for $\tilde \lambda$ at the 2.5PN order 
is simply given by Eq.(\ref{Eq53}) with $\tilde c_l (l)$
replaced by the expression for with $\tilde c_{\lambda} (l)$,
given by Eq. (\ref{Eq50d}).
This is so since contributions to $\tilde \lambda (l)$ arising 
from the  periastron advance constant $k$ appears at 
${\cal {O}}(c^{-7})$.
   In the next subsection, we plot analytic results obtained 
in these subsections and their influences on $h_\times$ and
$h_+$.

\subsection{ Graphical representation of the results}

    We begin the subsection by illustrating 
the temporal evolution of
$\bar c_\alpha$ and $ \tilde c_\alpha$. Next, we show the combined 
effects of these secular and periodic variations on basic angular
 variables that appear in the expressions for $h_\times$ and
$h_+$. Finally, we exhibit $h_\times$ and
$h_+$ evolving under gravitational radiation reaction
 and point out various features associated with post-Newtonian
accurate orbital motion.
In these figures, we terminate the orbital evolution 
when $j^2 = 48$. This criterion, as explained earlier, is chosen to
make sure that the orbit under investigation is a shrinking
slowly precessing ellipse.
We have also taken advantage of the `scaling' nature of the problem to
plot only dimensionless quantities in terms of dimensionless variables.
The conversion to familiar quantities like orbital frequency
$f$ (in Hertz) is given by $f \equiv \frac{n}{2\,\pi} = \frac{1}{2\,\pi}
\,\frac{c^3\, \xi}{G\,m} =  3.2312 \times 10^4\, \xi\,\frac{M_{\odot} }{m} $.
This implies that for a compact binary with $m= M_{\odot}$ and 
$\xi= 10^{-3} $, the orbital frequency will be $\sim 30 $ Hertz.

\begin{figure}
\resizebox{16cm}{!}{\includegraphics{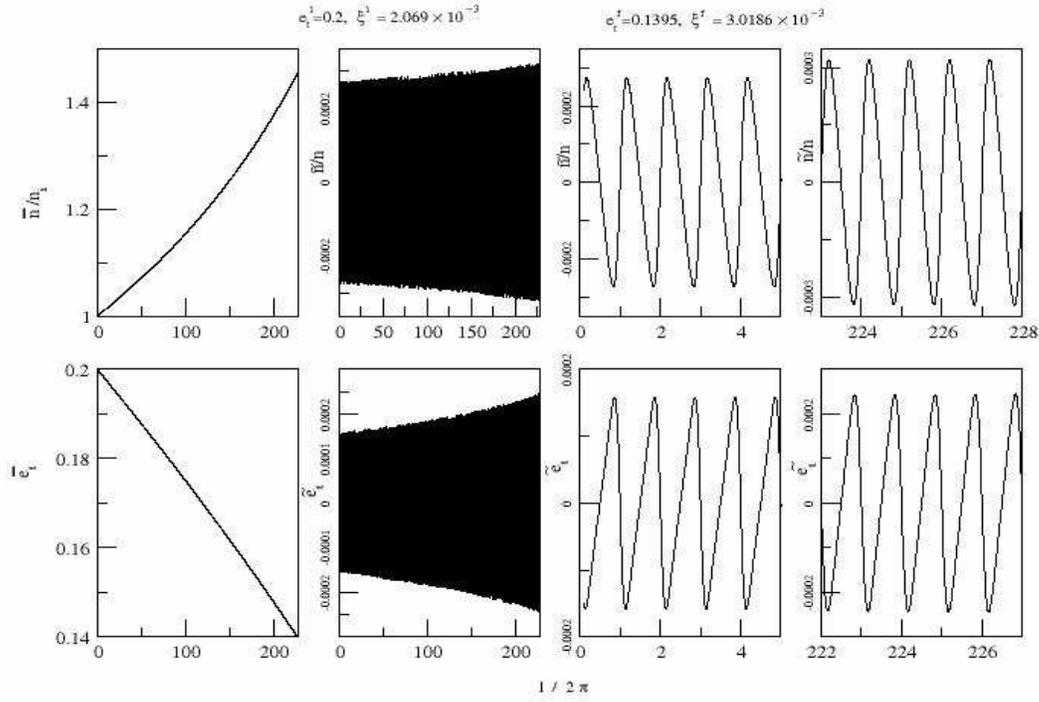}}
\caption{
The plots for $\bar n/n_i, \tilde n /n, \bar e_t$ and $\tilde e_t$
versus  $\frac{l}{2\,\pi}$, which gives the number of
orbital revolutions.
These variations are governed by the reactive 2.5PN equations of
motion. Periodic nature of the variations in $\tilde n$ and $\tilde e_t$
are clearly visible. $e_t^i$ and $e_t^f$ denote initial and final values for
the time eccentricity $e_t$, while $\xi^i$ and $\xi^f$ stand for similar 
values of  the adimensional mean motion $\frac{G\,m\, n}{c^3}$.
The plots are for $\eta=0.25$ and
the evolution is terminated when $j^2=48$. 
\label{fig:n_e_BT}
}
\end{figure}

\begin{figure}
\resizebox{16cm}{!}{\includegraphics{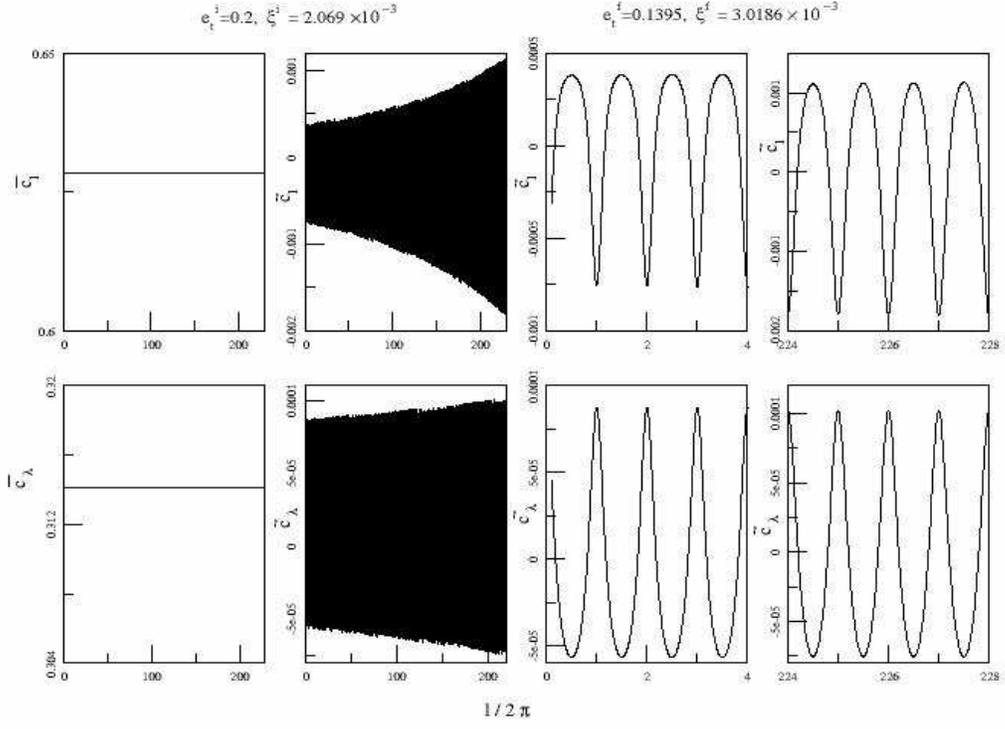}}
\caption {
The plots  $\bar c_l,\tilde c_l, \bar c_\lambda$ and $\tilde c_\lambda $
against orbital cycles, given by $\frac{l}{2\,\pi}$.
Similar to Fig. \ref{fig:n_e_BT},
these variations are governed by the reactive 2.5PN equations of
motion. Periodic nature of the variations in $\tilde c_l$ and 
$\tilde c_\lambda$ as well as the constancy of 
$\bar c_l$ and $\bar c_\lambda$ are clearly visible.
The symbols have the same meaning as in Fig \ref{fig:n_e_BT}.
}
\label{fig:cl_cm_BT}
\end{figure}

\begin{figure}
\resizebox{16cm}{!}{\includegraphics{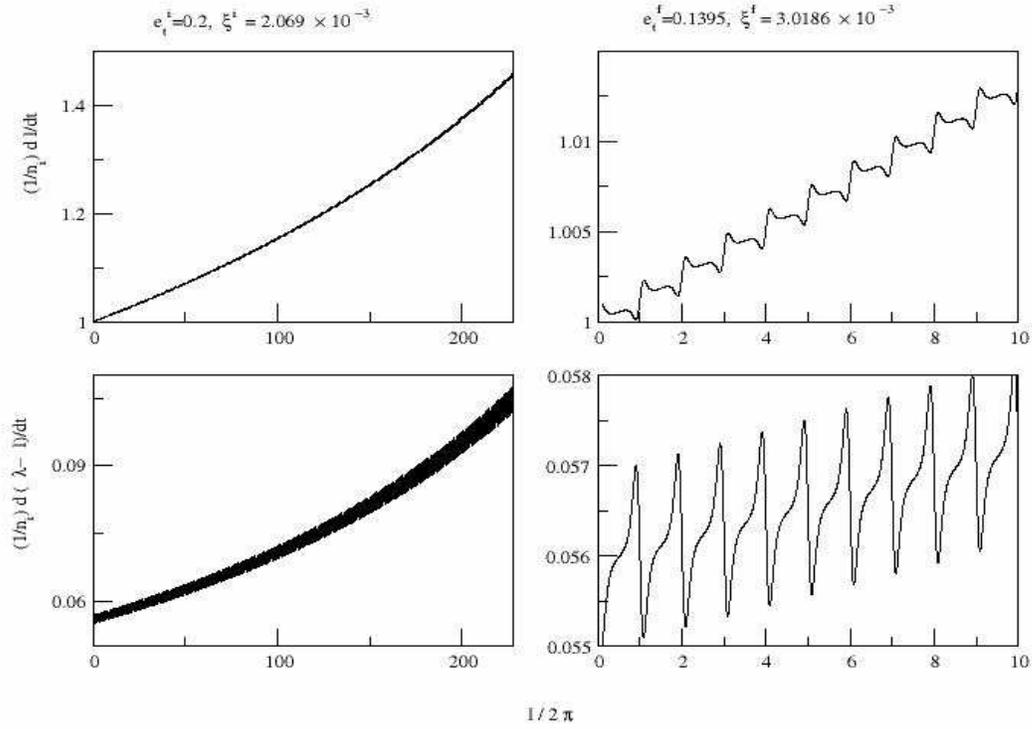}}
\caption {
The plots showing scaled time derivative of $l$ and
$\lambda -l $ as function of 
orbital cycles, given by $\frac{l}{2\,\pi}$.
The right panels show clearly both the secular drift and the
periodic oscillations of the plotted quantities.
Similar to earlier figures,
these variations are governed by the reactive 2.5PN equations of
motion. The initial and final values of the relevant orbital elements
are marked on the plots.
The plots are for $\eta=0.25$ and 
$n_i$ is the initial value of the mean motion $n$.
}
\label{fig:l_Lam}
\end{figure}

\begin{figure}
\resizebox{16cm}{!}{\includegraphics{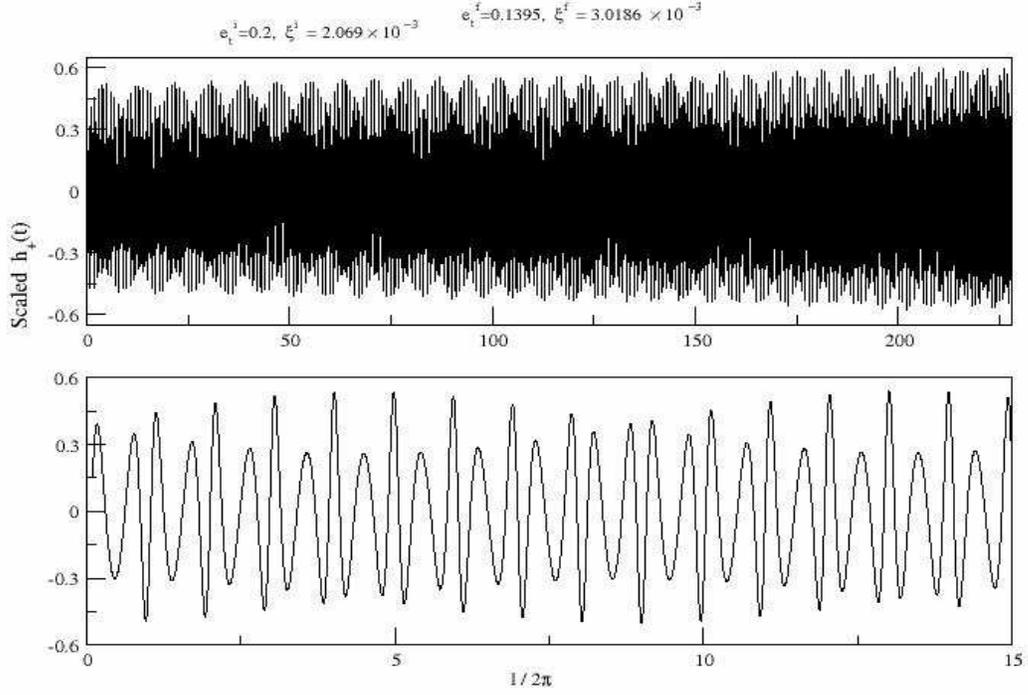}}
\caption {
The scaled $h_+$ ( Newtonian in amplitude and
2.5PN in orbital motion) 
plotted against orbital cycles, given by $\frac{l}{2\,\pi}$.
The `chirping', amplitude modulation due to periastron precession
are clearly visible in the upper panel.
In the bottom panel, we zoom into the initial stages of
orbital evolution to
show the effect of 
the periodic orbital motion  and the periastron advance on the
scaled $h_+(t)$.
The initial and final values of the relevant orbital elements
are marked on the plots.
The plots are for $\eta=0.25$ and the orbital
inclination is $i= \frac{\pi}{3}$.
}
\label{fig:P_P2}
\end{figure}

\begin{figure}
\resizebox{16cm}{!}{\includegraphics{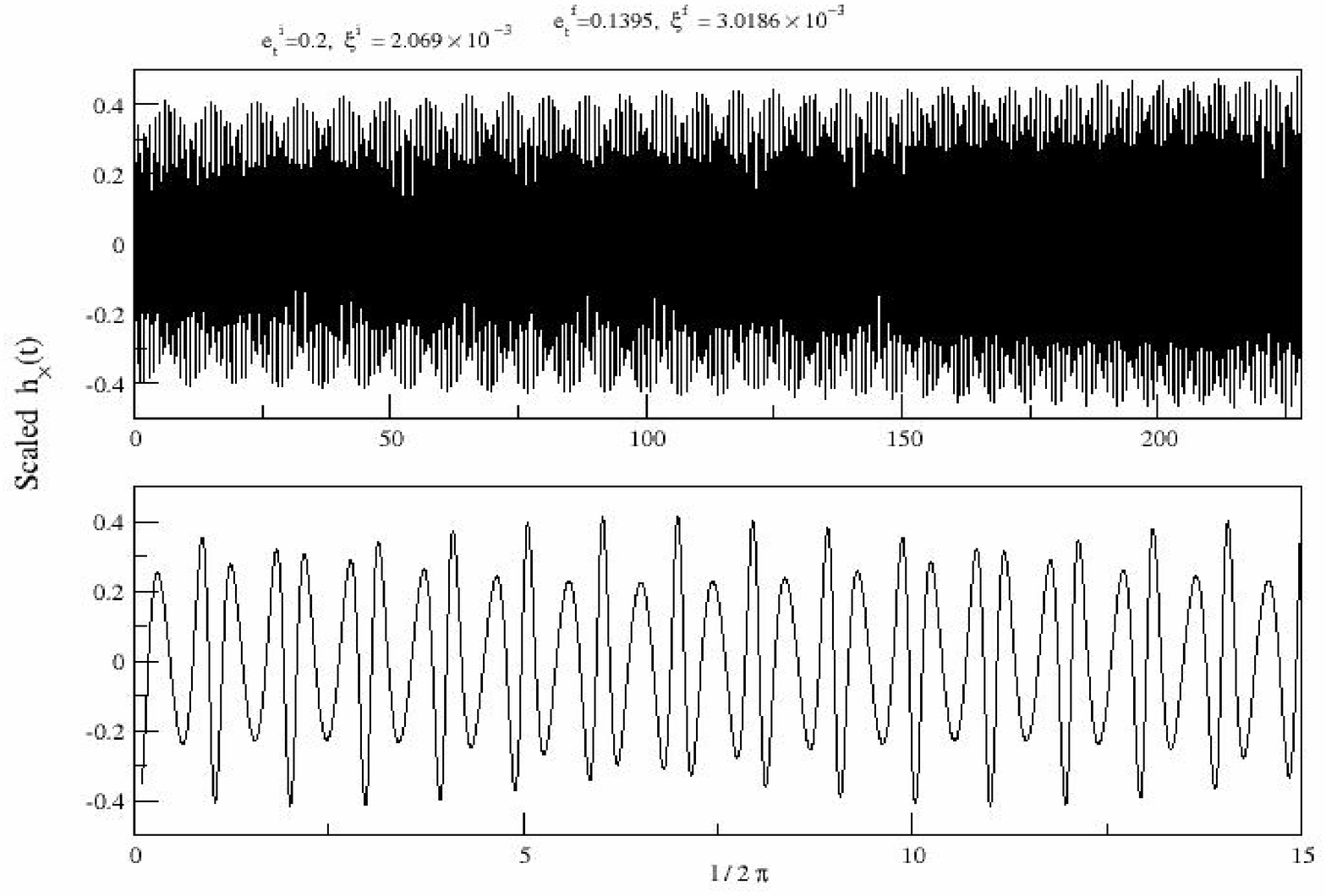}}
\caption {
The plots for the scaled $h_{\times}$ (Newtonian in amplitude and
2.5PN in orbital motion) as function of 
$\frac{l}{2\,\pi}$, where $l$ is
the mean anomaly .
The slow `chirping', amplitude modulation due to periastron precession
are clearly visible in the upper panel.
In the bottom panel, we zoom into the initial stages of
orbital evolution to
show the effect of
the periodic orbital motion  and the periastron advance on the
scaled $h_\times (t)$.
The initial and final values of the relevant orbital elements
are marked on the plots.
The plots are for a binary consisting of equal masses, such that 
$\eta=0.25$ and the orbital
inclination is $i= \frac{\pi}{3}$.
The orbital evolution, as in earlier cases, is terminated when 
$j^2 = 48$.
}
\label{fig:C_P2}
\end{figure}

   In Figs. (\ref{fig:n_e_BT}) and  (\ref{fig:cl_cm_BT}),
we plot $\bar n/n_i$ (where $n_i$ is the initial value of $n$), $\tilde n/ n,
\bar e_t, \tilde e_t, \bar c_l, \tilde c_l $,
$\bar c_\lambda$ and $ \tilde c_\lambda$ as  functions of
$\frac{l}{2\,\pi}$, which gives evolution in terms of elapsed orbital cycles.
We clearly see an  adiabatic increase (decrease) of $\bar n$ $(\bar e_t)$
as well as the periodic variations of $\tilde n$ and $ \tilde e_t$.
As expected, we also observe no secular evolution for $\bar c_l$ and
$\bar c_\lambda$, but clearly see periodic variations 
in  $\tilde c_l$ and $\tilde c_\lambda$.
In order to illustrate the effect of the secular and periodic
variations in the above orbital variables on the basic angles
$l$ and $\lambda$, in Fig. (\ref{fig:l_Lam}), we plot scaled 
$ dl/dt$ and $ d (\lambda -l )/dt$ as functions  of 
$\frac{l}{2\,\pi}$.
The figure  shows periodic oscillations superposed on the slow
secular drift.
Finally, in Figs. (\ref{fig:P_P2}) and (\ref{fig:C_P2}), we plot
scaled $h_+$ and $h_\times$ as  functions of $\frac{l}{2\,\pi}$.
We employ, for these figures, polarization amplitudes, which  are
Newtonian accurate while the orbital motion is 2.5PN accurate.
We clearly see `chirping' due to radiation damping, amplitude modulation 
due to periastron precession and also orbital period variations.

   Using the scaling argument mentioned earlier, we note that Figs. 
(\ref{fig:n_e_BT})-(\ref{fig:C_P2}) may be used to illustrate
the various aspects of a  compact binary inspiral from sources
relevant for {\em both} LIGO and LISA.
For instance, if we choose $m= 2.8 M_{\odot}$,  
the variation of $\xi$  from $2.069 \times 10^{-3}$ to
$3.0186 \times 10^{-3}$ in around $227$ orbital cycles corresponds to
orbital frequency variation from $150$ Hz to $\sim 217$ Hz in 
$\sim 8.15$ seconds.
Similarly,  the choice $m= 10^5 M_{\odot}$
corresponds to a binary inspiral involving two supermassive
black holes, where orbital frequency increases from $\sim 4.2 \times 10^{-3} $
Hz to $\sim 6.2 \times 10^{-3}$ Hz in $\sim 2.7$ days.

  Finally, we note that  Figs. (\ref{fig:n_e_BT})-(\ref{fig:C_P2}), 
were drawn mainly to exhibit {\em more clearly}
the existence of {\it periodic} variations in orbital
elements (analytically investigated for the first time in the present
paper) due to the radiation reaction.
Evidently, as these variations scale as $\xi^{5/3}$, see Eqs.(\ref{Eq50}),
they become quite small if the binary is not near the plunge boundary.
Our work is important, even when the binary is not near the  LSO,
as it shows, in a technically clear manner, how to describe
the exact phasing as the sum of the usually considered 
`adiabatic' phasing (involving only secular variations)
and a   normally neglected `post-adiabatic'
phasing (involving only the relatively smaller `periodic' variations).
More about this in the conclusions below.

\section{PN accurate  adiabatic evolution for $\bar n$ and $\bar e_t$}
\label{PNadiabaticSec}

  One of the useful results of the present work are  Eqs. (\ref{Eq43})
and (\ref{Eq44}).
Indeed, these equations provide a clear justification for the 
usually considered `adiabatic' approximation (in cases, 
where one is sufficiently away
from the plunge boundary so that one can safely neglect the additional
periodic contributions and treat orbits to be quasi-Keplerian).
More precisely, we have,  earlier, proved that Eqs. (\ref{Eq43})
would be valid, even if we are considering radiation reaction to
the accuracy ${\cal O}(c^{-5}) + {\cal O}(c^{-7}) +{\cal O}(c^{-8}) 
+{\cal O}(c^{-9})$.
Concerning Eqs. (\ref{Eq44}), 
following its derivation again, we see that (if we were to define
$c_a$ to sufficient accuracy, by considering the conserved 
quantities of the `conservative part' of the dynamics)
it is valid up to terms which are quadratic in the  radiation reaction,
i.e up to ${\cal O}(c^{-10})$.
Therefore, to those very high accuracies, our work shows that
the secular part of the phasing (in the sense of the decomposition,
as in Eq.~(\ref{Eq25}) ) is essentially given by the simple averaged
result, given by Eqs. (\ref{Eq43}) and (\ref{Eq44}).
Then, as usual, we expect that the averaged losses of the 
{\it mechanical} energy and angular momentum of the system,
appearing in Eqs.~(\ref{Eq44}), are equal to the corresponding
{\it far zone} (FZ) fluxes of energy and angular momentum
in the form of radiated gravitational waves.

  To complete this work, let us briefly show how to obtain,
in ADM coordinates, the 2PN accurate secular changes in 
$\bar n$ and $\bar e_t$, equivalent to corresponding
2PN accurate FZ fluxes of energy and angular momentum.
The differential equations for 
 $\bar n$ and $\bar e_t$
are computed, following the 
PN accurate calculations presented in
\cite{GS89}.
These computations require PN corrections to orbital averaged
expressions for the far-zone energy and angular momentum fluxes
and PN accurate expressions for $n$ and $e_t$, all of these
expressed in terms of orbital energy and angular momentum.
The PN accurate expressions for $\frac{d\bar n}{dt}$
and $\frac{d\bar e_t}{dt}$
are obtained by differentiating PN accurate 
expressions for $ n$ and $ e_t$
w.r.t. time and then heuristically equating
the time derivatives of orbital energy and angular momentum
to orbital averaged
expressions for the far-zone energy and angular momentum fluxes.
For the ease of implementation, we split 2PN accurate computations of 
$\frac{d\bar n}{dt}$ and $\frac{d\bar e_t}{dt}$
into two parts.
The first part deals with the purely `instantaneous' 2PN corrections 
and the second part considers the so called `tail' contributions,
appearing at the 1.5PN (reactive) order
\cite{Def_Inst}.
The computations to get `instantaneous' contributions
begin with 2PN corrections to far-zone fluxes, in harmonic
gauge, in terms of $r, \dot r $ and $ v^2$ available in
\cite{GI97}.
Using  2PN accurate relations connecting the dynamical
variables in harmonic and ADM coordinates, given 
by Eqs.~(\ref{ctad}) in Appendix~\ref{AppA},
we obtain expressions for the far-zone fluxes in ADM coordinates.
These far-zone fluxes are orbital averaged, using 2PN accurate
 generalized quasi-Keplerian parametrization for elliptical
orbits, following lower PN computations done in \cite{GS89}.
We then compute time derivative of PN accurate expressions for $n$
and $e_t$ and equate resulting
time derivatives of orbital energy and angular momentum to
orbital averaged
expressions for the far-zone energy and angular momentum fluxes
respectively, to get PN accurate  $\frac{d \bar n}{dt}$ and 
$\frac{d \bar e_t }{dt}$
in terms of $E,j, m$ and $\eta$. Finally, we use Eqs. (\ref{inp2})
to obtain the differential equations for $\bar n$ and $\bar e_t$
in terms of $ n, e_t, m$ and $\eta$.
The tail contributions to $\frac{d \bar n}{dt}$ and
$\frac{d \bar e_t }{dt}$ are already available, in  slightly different
forms, in \cite{GS89} and we only rewrite these expressions in 
our $n$ and $e_t$ variables.
Adding these `instantaneous' and `tail' contributions gives
us the following 2PN accurate evolution equations for 
$\bar n$ and $\bar e_t$.
\bs
\bea
\frac{d\bar n}{dt} &=&
\frac{ (G\,m\,n)^{11/3}\,\eta }{G^2\,m^2\,c^5}
\biggl \{
{\dot {\bar n}}^{{\rm N}} + 
{\dot {\bar n}}^{1\,{\rm PN}} + 
{\dot {\bar n}}^{1.5\, {\rm PN}} + 
{\dot {\bar n}}^{2\, {\rm PN}}  
\biggr \}\,,
\\
\frac{d\bar e_t}{dt} &=&
-\frac{ (G\,m\,n)^{8/3}\,\eta\,e_t}{G\,m\,c^5}
\biggl \{
{\dot {\bar {e_t} }}^{{ {\rm N}}} + 
{\dot {\bar {e_t} }}^{1\, {\rm PN}} + 
{\dot {\bar {e_t} }}^{1.5\, {\rm PN} } + 
{\dot {\bar {e_t} }}^{2\, {\rm  PN}} \,, 
\biggr \}
\eea
\es
where  
${\dot {\bar n}}^{{\rm N}}, {\dot {\bar n}}^{1\,{\rm PN} } 
{\dot {\bar n}}^{2\,{\rm PN}},  {\dot {\bar {e_t} }}^{{\rm N}}, 
{\dot {\bar {e_t} }}^{1\,{\rm PN}}$ 
and $ {\dot {\bar {e_t} }}^{2\,{\rm PN}}$  
denote  `instantaneous' contributions 
to 2PN order, while 
$ {\dot {\bar n}}^{1.5\,{\rm PN}} $ and
$ {\dot {\bar {e_t} }}^{1.5\,{\rm PN}}$ stand for
the `tail' contributions. 
Using 2PN accurate orbital representation and far-zone energy and angular
momentum fluxes, 
we have computed 2PN accurate `instantaneous' contributions
to $\frac{d\bar n}{dt} $
and $\frac{d\bar e_t}{dt} $,
whose explicit forms
are given by
\bs
\bea
{\dot {\bar n}}^{{\rm N} } &=& 
\biggl \{ \frac{1}{5\, (1-e_t^2)^{7/2} }
\biggl [ 96\, + 292\,e_t^2 + 37\,e_t^4 \biggl ]
\biggl \}\,,
\\
{\dot {\bar n}}^{1\,{\rm PN}} &=& 
 \frac{ \xi^{2/3} }{ 280\, (1-e_t^2)^{9/2} }
\biggl \{ 
20368 - 14784\,\eta + \left ( 219880 - 159600\,\eta \right )\,e_t^2
\no
&&
+ \left ( 197022 - 141708\,\eta \right )\,e_t^4
+ \left ( 11717 -8288\, \eta \right )\,e_t^6
\biggr \} \,,
\\
{\dot {\bar n}}^{2\,{\rm PN} } &=& 
\frac { \xi^{4/3} }{ 30240\,(1-e_t^2)^{11/2}}
\biggl \{
 \biggl ( 
12592864 - 13677408\,\eta + 1903104\,{\eta}^{2}
\biggr )
\no
&&
+\biggl ( 
131150624 -217822752\,\eta + 61282032\,{\eta}^{2}
\biggr )\,e_t^2
\no
&&
+ \biggl ( 
282065448 - 453224808\,\eta + 166506060\,\eta^2
\biggr )\,e_t^4
\no
&&
+ \biggl ( 
112430610 -144942210\,\eta + 64828848\,{\eta}^{2}
\biggr )\,e_t^6
\no
&&
+ \biggl ( 
3523113 -3259980\,\eta + 1964256\,{\eta}^{2}  \biggr )\,e_t^8
\no
&&
-3024\, \biggl (
96 +4268\,e_t^{2} +4386\,e_t^{4}
+ 175\,e_t^{6}
\biggr )  \left( 2\,\eta-5
 \right) {\sqrt {1-e_t^2} }
 \biggr \}\,,
\\
{\dot {\bar {e_t} }}^{{\rm N} } &=&
\biggl \{
\frac{1}{15\,(1-e_t^2)^{5/2}}
\biggl [ 304 + 121\,e_t^2 \biggr ]
\biggr \}\,,
\\
{\dot {\bar {e_t} }}^{1\,{\rm PN}} &=&
\frac{\xi^{2/3} }{2520\,(1-e_t^2)^{7/2}}
\biggl \{
340968 - 228704\,\eta + \left ( 880632 - 651252\,\eta \right )\,e_t^2
\no
&&
+ \left ( 125361 -93184 \,\eta \right )\,e_t^4
\biggr \}\,,
\\
{\dot {\bar {e_t} }}^{2\,{\rm PN}} &=&
\frac {\xi^{4/3} }{30240\, (1-e_t^2)^{9/2} }
\biggl \{
20621680
-28665360\,\eta
+ 4548096\,{\eta}^{2}
\no
&&
+ \biggl (
86398044\,
-148804812\,\eta
+ 48711348\,{\eta}^{2}
\biggr )\,e_t^2
\no
&&
+ \biggl (
69781286\,
-95827362\,\eta+
42810096\,{\eta}^{2}
\biggr )\,e_t^4
\no
&&
+ \biggl (
3786543\,
-4344852\,\eta+
2758560\,{\eta}^{2}
\biggr )\,e_t^6
\no
&&
-1008\,
 \biggl (
  2672 +  6963\,e_t^2 + 565\,e_t^4
   \biggr )
     \left( 2\,\eta-5 \right) \sqrt {1-e_t^2}
      \biggr \}\,,
 \eea
 \es
where, as in earlier instances, $n$ and $e_t$ on the right hand side
of these equations stand for $\bar n$ and $\bar e_t$.

    The `tail' contributions to $\frac{d\bar n}{dt} $ 
and $\frac{d\bar e_t}{dt} $, which appear at 1.5PN order,
are derivable 
using Keplerian orbital parameterization
and 
`tail' corrections to orbital averaged expressions for the far-zone
energy and angular momentum fluxes available in \cite{BS93,RS97}.
For the ease of presentation, we display below `tail' contributions to
$\frac{d\bar n}{dt} $
and $\frac{d {\bar e_t}^2}{dt} $, rather than 
$\frac{d\bar n}{dt} $
and $\frac{d\bar e_t}{dt} $
\bs
\bea
\biggl ( \frac{d\bar n}{dt} \biggr ) {\bigg |}_{\rm Tail} &=& \frac{384}{5}\,
\frac{ (G\,m\,n)^{(14/3)}\,\pi\, \eta}{G^2\,m^2\,c^8}
\, \kappa_{\rm E}\,,
\\
\biggl (\frac{d {\bar e_t}^2 }{dt} \biggr ) {\bigg |}_{\rm Tail} &=&
-\frac{256}{5}\,\frac{ (G\,m\,n)^{(11/3)}\,\pi\,\eta}{G\,m\,c^8}
\biggl \{ (1-e_t^2)\,\kappa_{\rm E} 
- \sqrt { 1-e_t^2}\, \kappa_{\rm J}
\biggr \}
\,,
\eea
\es
where both $\kappa_{\rm E}$ and $\kappa_{\rm J}$
are expressible
in terms of infinite sums involving 
quadratic products of Bessel functions $J_p (p\,e_t)$
and its derivative $J'_p (p\,e_t)$.
For completeness,
the explicit expressions for $\kappa_{\rm E}$ and $\kappa_{\rm J} $,
given in \cite{BS93,RS97}, are listed below
\bs
\bea
\kappa_{\rm E} &=&
\sum_{p=1}^{+\infty} \frac{p^3}{4}
\biggl \{ ( J_p (p\,e_t))^2 
\biggl [ \frac{1}{e_t^4} - \frac{1}{e_t^2} + \frac{1}{3}
+ p^2 \biggl ( \frac{1}{e_t^4} - \frac{3}{e_t^2} + 3 -e_t^2
\biggr ) \biggr ]
\no
&&
+ p\, 
\biggl [ -\frac{4}{e_t^3} + \frac{7}{e_t} -3\,e_t \biggr ]
J_p (p\,e_t)\,J'_p (p\,e_t)
+ ( J'_p (p\,e_t) )^2 
\biggl [ \frac{1}{e_t^2} - 1 
\no
&&
+ p^2 \left ( 
\frac{1}{e_t^2} -2 + e_t^2 \right )
\biggr ]
\biggr \}\,,
\\
\kappa_{\rm J} &=&
\sum_{p=1}^{+\infty} \frac{p^2}{2} \sqrt{ 1-e_t^2}
\biggl \{
p\, \biggl [ \frac{3}{e_t^2} -\frac{2}{e_t^4} -1 \biggr ]
\,( J_p (p\,e_t))^2
+ \biggl [ \frac{2}{e_t^3} -\frac{1}{e_t}
\no
&&
+ 2\,p^2 \biggl ( \frac{1}{e_t^3} -\frac{2}{e_t} + e_t
\biggr ) \biggr ] 
 J_p (p\,e_t)\,J'_p (p\,e_t)
 + 2\,p \biggl ( 1 -\frac{1}{e_t^2} \biggr )\,
( J'_p (p\,e_t) )^2
\biggr \}\,.
\eea
\es
We have checked that to 1PN order the  above equations are 
consistent with expressions for $\frac{d \bar a_r}{dt}$
and $\frac{d \bar e_r}{dt}$, computed in \cite{GS89}.
At 2PN order, the  above expressions are also  consistent with 
{\em corrected}  formulae for $\frac{d \bar a_r}{dt}$
and $\frac{d \bar e_r}{dt}$, available in \cite{GI97,GI03er}.

\section{ Conclusion}
\label{ConcludeS}
Let us summarize what we have proposed in this paper and 
point out possible extensions.
We have provided a method for analytically constructing high-accuracy
templates for the gravitational wave signals emitted by compact 
binaries when they move in inspiralling 
slowly precessing
eccentric orbits.
In  contrast to the simpler problem of modeling the gravitational
wave signals emitted by inspiralling {\it circular} orbits,
which contain only two different time scales (orbital period and
radiation reaction time scale), the case of {\it inspiralling eccentric}
orbits involves {\it three different time scales}: orbital period,
periastron precession and radiation-reaction time scales.
An improved `method of variation of constants' is used 
to combine these three time scales, without making the usual
approximation of treating 
adiabatically the radiative time scale. 
By going to a suitable center-of-mass frame, the transverse
traceless (TT) radiation field and hence the GW polarizations are
expressed as PN expansions of the form given in Eq. (\ref{Eq4})
in harmonic coordinates involving only the relative position and
 velocity.
The polarisations can be rewritten in terms of the
ADM positions and velocities by using the contact transformations
available to move from the harmonic  coordinates  to the ADM
coordinates.
In the ADM coordinates, the unperturbed 2PN (3PN) motion may
be explicitly solved by a  generalized quasi-Keplerian representation
involving two angles $l$ and $\lambda$ and four constants of the 2PN (3PN)
motion $c_\alpha= c_1,c_2,c_l$ and $c_\lambda$.
By a Lagrange method of variation of constants, this unperturbed
solution is used to
prescribe
a general solution to the perturbed
(by radiation-reaction) system of the form given by
Eqs. (\ref{Eq17}) and (\ref{Eq18}),
in terms of the varying $c_\alpha$'s.
Among the four $c_\alpha$'s, two ($c_l$ and $c_\lambda$)
are found to be constants, while
the other  two 
$c_\alpha$'s
satisfy two coupled first order differential equations.
A two scale decomposition of the $c_\alpha$'s is made to model
the combination of the slow (radiation reaction, secular) drift
and the fast (orbital time scale, periodic) oscillations.
This allows us to decompose the TT gravitational wave amplitudes
or polarizations into a part associated with the secular variations
and another part 
associated with  the fast oscillations.
If one 
expands in the fast oscillations,
the  oscillatory contributions to the phasing 
would be  equivalent to new additional
 ${\cal{O}}(v^5/c^5)$ contributions
(which stay always small )
to be added to the usually considered 2.5PN  
amplitude $h_{ij}^5$ in the adiabatic approximation.
Therefore, as long as an amplitude correction ${\cal{O}}(v^5/c^5)$ is
not needed our results show that it is enough to use only secularly
varying $\bar{n}$ and $\bar{e}_t$.
Note that our description of secular effects automatically include
an effect of a secular acceleration of the periastron precession,
analogous to the usual secular acceleration of the orbital motion.
Namely, the secular part of the angle $\lambda -l$ measuring
the periastron longitude varies as
\be
\bar{\lambda}-\bar{l}=\int \bar{k}\bar{n}dt
\ee
where $\bar{k}(t)=\delta\Phi/2\pi$ is the secularly varying periastron
precession per orbital period.

Schematically,  thus our work may be summarized as follows:
\vspace*{0.3in}
\newline
\framebox(280,25)[c]{\Large $ h_{+,\times}(r^{\rm harm},\phi^{\rm harm},
\dot{r}^{\rm harm},\dot{\phi}^{\rm harm})$ } 
\newline
\mbox{ \hspace{1.5in} \Large $\Downarrow$}
\vspace*{0.1in}
\newline
\framebox(280,25)[c]{ \large
$h_{+,\times}(r^{\rm ADM},\phi^{\rm ADM},\dot{r}^{\rm ADM},
\dot{\phi}^{\rm ADM})$}
\newline
\mbox{ \hspace{1.5in} \Large $\Downarrow$}
\vspace*{0.1in}
\newline
\framebox(280,25)[c]{ \large
$h_{+,\times}(n,e_t,l,\lambda) $}
\newline
\mbox{ \hspace{1.5in} \Large $\Downarrow$}
\vspace*{0.1in}
\newline
\framebox(280,25)[c]{ \large
$ h_{+,\times}(\bar{n}+\tilde{n},\bar{e}_t+\tilde{e}_t,
\bar{l}+\tilde{l},\bar{\lambda}+\tilde{\lambda}) $}
\newline
\mbox{ \hspace{1.5in} \Large $\Downarrow$}
\vspace*{0.1in}
\newline
\framebox(350,40)[c]{ \large
$h_{+,\times}(\bar{n},\bar{e}_t,
\bar{l},\bar{\lambda})+
\left\{\frac{\partial h_{+,\times}}{\partial n} \tilde{n}+
\frac{\partial h_{+,\times}}{\partial e_t} \tilde{e_t}+
\frac{\partial h_{+,\times}}{\partial l} \tilde{l}+
\frac{\partial h_{+,\times}}{\partial \lambda} \tilde{\lambda}\right\} $}
\newline
\hspace*{1.8in}
\mbox{  \Huge $\equiv$}
\vspace*{0.1in}
\newline
\vspace*{0.2in}
\hspace*{0.2in}
\framebox(280,40)[c]{ \Large
$
h_{+,\times}(\bar{n},\bar{e}_t, \bar{l},\bar{\lambda})+
\frac{1}{c^5}\,\delta 
h_{+,\times}^5(\tilde{n},\tilde{e}_t, \tilde{l},\tilde{\lambda})
$}

      In this paper, the orbit of the binary was treated to be an inspiralling,
slowly precessing ellipse, which prevented us from approaching
the LSO. However, as mentioned earlier, using the EOB approach, we intend to
explore the orbital dynamics and hence the evolution of
gravitational wave polarizations near the  LSO in the near future.
There are also quite a few generalizations which can be tackled using the
formalism presented here and we list a few of them here. 
In this paper,  
the conservative dynamics was restricted
to the 2PN order and therefore,
a natural extension will be to incorporate
 the 3PN conservative
dynamics.  We also restricted our approach
to compact binaries consisting of non-spinning point masses.
It is possible to extend the generalized quasi-Keplerian
parametrization and hence our method to spinning compact binaries.
To do that, first, one needs to extend the quasi-Keplerian
representation to include the  effects due to spin-orbit and
spin-spin interactions, which requires 
generalizing a restricted analysis done in
 \cite{W95} and this is under investigation \cite{KG04}.
Finally,  starting from    the  gravitational wave polarizations
from {\it inspiralling eccentric} binaries in the time-domain,
it should be possible to  perform  a  spectral analysis  and
 see how their power spectrum depends on various orbital elements like
$n$, $e_t$ and $i$.

\begin{acknowledgments}

It is a pleasure to thank Gerhard Sch\"afer for discussions.
BRI thanks IHES for hospitality during different stages of
the work.
AG gratefully acknowledges the financial support of
the Deutsche
Forschungsgemeinschaft (DFG) through SFB/TR7
``Gravitationswellenastronomie'' and 
the Natural Sciences and Engineering Research
Council of Canada. In addition, AG  thanks 
Bernie Nickel, Eric Poisson and 
Gerhard Sch\"afer for encouragements 
and appreciates  the hospitality of IHES during the
final stage of this work.
\end{acknowledgments}

\appendix

\section{The construction of PN corrections to $h_\times$ and $h_+$}
\label{AppA}

   In this appendix, we sketch the procedure to compute PN corrections
to
 $h_+$ and $h_\times$
in ADM coordinates, in terms of the dynamical
variables $r, \dot r, \phi$ and $\dot \phi$.
It is clear from Eqs. (\ref{Eq1}), (\ref{Eq3}) and (\ref{Eq6}) that 
PN corrections $h_+$ and $h_\times$ 
require PN corrections to $h_{ij}^{\rm TT}$.
The  `instantaneous'
2PN accurate contributions to $h_{ij}^{\rm TT}$
in harmonic coordinates in terms of 
the components of ${\bf n} $ and ${\bf v} $,
$ r, \dot r$ and $v^2$
were computed in \cite{GI97}.
To obtain similar expressions in ADM coordinates, we employ 
the 2PN accurate contact transformations linking the harmonic
and ADM coordinates, given in \cite{DS85}, which prescribe
the way to relate the dynamical variables in these coordinates.
We list below the transformation equations
relating ${\bf x}$ and ${\bf v}$ in the harmonic coordinates
to the corresponding ones in 
ADM \cite{GI03er}:
\bs
\label{ctad}
\bea
{\bf x}_{ \rm H} &=& {\bf x}_{ \rm A}
+ { Gm \over 8\,c^4\,r}\left \{ \left [ \left ( 5 v^2
-\dot r^2 \right )\eta +  2 \,\left ( 1
+12\eta \right ){ Gm\over r} \right ] { \bf x}
\right.\nonumber\\
& & \left.
-  18 \,\eta \, r\dot r\,{\bf v} \right \}\,,\\
{ \bf v }_{ \rm H} &=& { \bf v }_{ \rm A} - { Gm \dot r\over 8\,c^4\, r^2}
\biggl \{ \biggl [ 7 v^2 + 38 { Gm \over r}
- 3\dot r^2 \biggr ]\eta + 4\,{ G m \over r}\, \biggr \}
{ \bf x}
\nonumber \\
& &
-{ Gm \over 8\,c^4 r} \biggl \{
\biggl [ 13\, v^2 - 17 \dot r^2 - 42 { G m\over r} \biggr ]\eta
- 2\,{ Gm \over r} \biggr \}{ \bf v} \,.
\eea
\es
The subscripts `${\rm H}$' and `${\rm A}$' denote
quantities in the  harmonic  and
in the ADM coordinates respectively.
Note that in all the above  equations the differences between the
two gauges are at the  2PN order
and hence no suffix is  used 
for the 2PN terms.
We do not list the transformation relations for $r, \dot r $ and $v^2$
as they easily follow from Eqs. (\ref{ctad}), as
$r = |{\bf x}|, r\,\dot r = {\bf x}\cdot {\bf v}
$ and $v^2= {\bf v}\cdot {\bf v} = \dot r^2 + r^2\,\dot \phi^2 $.

  Using Eqs.(\ref{ctad}),
the 2PN corrections to $ h_{ij}^{TT}$ in ADM coordinates
can easily be obtained from
Eqs. (5.3) and (5.4) of \cite{GI97}.
For economy of presentation, we write
$ (h_{ij}^{TT})_{\rm A}$
in the following manner,
$ (h_{ij}^{TT})_{\rm A} =  (h_{ij}^{TT})_{\rm O} + $ `$ Corrections $', where
$(h_{ij}^{TT})_{\rm A}$ represent
the metric perturbations in the ADM coordinates.
$ (h_{ij}^{TT})_{\rm O}$ is a short hand notation for
expressions on the R.H.S of Eqs. (5.3) and (5.4) of \cite {GI97},
where $ {\bf N},{\bf n}, {\bf v}, v^2, \dot r, r $ are the ADM variables
${\bf N}_{\rm A},{\bf n}_{\rm A}, {\bf v}_{\rm A},v_{\rm A}^2,
{\dot r}_{\rm A},
r_{\rm A} $ respectively.
The `$ Corrections $' represent the differences at the 2PN
order, that arise due to the change of the coordinate system,
given by Eqs. (\ref{ctad}). As the two coordinates are
different only at the 2PN order, the `$ Corrections $' come
only from the leading Newtonian terms in Eqs. (5.3) and (5.4)
of \cite{GI97}.
\bea
\label{hijA}
(h^{TT}_{ij})^{ }_{\rm A} & =&
(h^{TT}_{ij})^{ }_{\rm O} +
{G \over c^4\,R}\,{ G\,m \over 2\,c^4\,r_{\rm A}} \biggl \{
\biggl [
(5 \,v_{\rm A}^2 -55\,{\dot r}_{\rm A}^2)\,\eta
+ 2\left ( 1 +12\,\eta \right )
\,{G\,m \over r_{\rm A}} \biggr ] \,{G\,m\over r_{\rm A}}\,(n_{ij})_{\rm A}^{TT}
\no
& &
- 2\,\biggl [
(7\,v_{\rm A}^2 -3\,{\dot r}_{\rm A}^2)\,\eta
+4\left ( 1
+5\,\eta \right )\,{G\,m \over r_{\rm A}} \biggr ]
{\dot r}_{\rm A}
\,(n_{(i}v_{j)})_{\rm A}^{TT}
\no
& &
- \biggl [
(26 \,v_{\rm A}^2 -34 \,{\dot r}_{\rm A}^2)\,\eta
-\left ( 4 + 84\,\eta \right )\,{ G\,m \over r_{\rm A}}\biggr ]
\,(v_{ij})_{\rm A}^{TT}
\biggr \}\,.
\eea

   We now have all the inputs required to compute the 2PN corrections to
$h_+$ and $h_\times$ in ADM coordinates, in terms of $r, \phi, \dot r$
and $\dot \phi$. We write, schematically, 
the expression for $(h^{TT}_{ij})^{ }_{\rm A}$ as
\be
\label{hijsym}
(h^{TT}_{ij})_{\rm A} = \alpha_{vv} \,v_{ij} +
\alpha_{nn}\,n_{ij}
+ \alpha_{nv} \, n_{(i}v_{j)}\,.
\ee
We apply exactly the same procedure which gave us the Newtonian 
expressions to $h_+$ and $h_\times$ from Newtonian
contributions to $h^{TT}_{ij}$.
Since the explicit expressions for the
final 2PN accurate `instantaneous' GW polarization states
are too lengthy to be listed here, we write 
schematically 
\bs
\bea
h_{+} &=& \biggl\{ \alpha_{vv}\biggl[(\cdots)
+ (\cdots) \cos 2\phi + (\cdots)\sin 2\phi\biggr]
+\alpha_{nn}\biggl[(\cdots) + (\cdots)\cos 2\phi\biggr]
\nonumber\\
&& + \alpha_{nv}\biggl[(\cdots)+(\cdots)\cos 2\phi
+(\cdots)\sin 2\phi\biggr]\biggl\}\,,\\
h_{\times} &=& \biggl\{ \alpha_{vv}\bigl[ (\cdots) \cos 2\phi
+ (\cdots)\sin 2\phi\bigr]
+\alpha_{nn}\biggl[ (\cdots)\sin 2\phi\biggr]
\no
&&
+
\alpha_{nv}\biggl[(\cdots)\cos 2\phi
+(\cdots)\sin 2\phi\biggr]\biggl\}.
\eea
\es
In the above and what follows
$\bigl(\cdots\bigr)$ denotes various coefficients
expressed in terms of $r, \dot r, \dot \phi, m_1, m_2 $
and $ i$. 
The structure of the PN expansion of the
 coefficients $\alpha_{ij}$ above
is the following:
\bea
 \alpha_{vv}&\sim& 1+
\frac{1}{c} \biggl[\bigl(\cdots\bigr)\cos \phi
+ \bigl(\cdots\bigr)\sin \phi\biggr]+
\frac{1}{c^2} \biggl[\bigl(\cdots\bigr)
+ \bigl(\cdots\bigr)\cos 2\phi
+ \bigl(\cdots\bigr)\sin 2\phi\biggr]+\nonumber\\
&&\frac{1}{c^3} \biggl[\bigl(\cdots\bigr)\cos 3\phi
+ \bigl(\cdots\bigr)\sin 3\phi\biggr]+
\frac{1}{c^4} \biggl[\bigl(\cdots\bigr)
+ \bigl(\cdots\bigr)\cos 4\phi + \bigl(\cdots\bigr)\sin 4\phi\biggr].
\eea
$\alpha_{nn}$
and $\alpha_{nv}$
have similar expansions with the exception
that for $\alpha_{nv}$ 
the leading order term is at $1\over c$ order. 
We note that, similar to Eq. (\ref{Eq6}), the above sketched
expressions for $h_+$ and $h_\times$ are in a form suitable to
apply our phasing formalism.

\newpage

\newpage

\newpage

\newpage

\newpage

\newpage

\begin{thebibliography}{999}


\bibitem{GWIFs}
Following URLs provide a wealth of information about 
the terrestrial gravitational wave interferometers.
\url{http://www.ligo.caltech.edu},
\url{http://www.virgo.infn.it},
\url{http://www.geo600.uni-hannover.de}, and
\url{http://tamago.mtk.nao.ac.jp}.

\bibitem{GLPPS}
L.\ P.\ Grishchuk, V.\ M.\ Lipunov, K.\ A.\ Postnov, M.\ E.\ Prokhorov,
and B.\ S.\ Sathyaprakash,
Phys.\ Usp.\ {\bf 44}, 1 (2001); Usp.\ Fiz.\ Nauk {\bf 171}, 3 (2001),
(astro-ph/0008481).
\bibitem{DIJS}
There are  a large  number of research articles
which investigate issues  related to the construction of
accurate and efficient `ready to use`
search templates for compact binaries of arbitrary mass ratio,
moving {\em mainly} in quasi-circular orbits. See e.g. 
T.\ Damour, B.\ R.\ Iyer and  B.\ S.\ Sathyaprakash,
Phys.\ Rev.\ D {\bf 63 }, 044023 (2001); {\it ibid} {\bf 66}, 027502 (2002)
and references therein.
Issues related to binary black holes are exhaustively dealt in 
A. Buonanno, Y. Chen and M. Vallisneri, Phys. Rev. D {\bf 67}, 024016 (2003)
and 
T.\ Damour, B.\ R.\ Iyer, P.\  Jaranowski and  B.\ S.\ Sathyaprakash
Phys.\ Rev.\ D {\bf 67}, 064028, (2003). 
\bibitem{P64}
P.\ C.\ Peters, Phys.\ Rev.\ {\bf 136}, B1224 (1964).
\bibitem{MH02}
M.\ C. \ Miller and  D.\ P.\  Hamilton, 
Astrophys.\ J.\, {\bf  576}, 894, (2002).

\bibitem{K62}
Y.\ Kozai, Astro.\ J.\, {\bf 67}, 591, (1962).

\bibitem{W03}
L.\ Wen,
{\it On the Eccentricity Distribution of Coalescing Black Hole Binaries Driven by the Kozai Mechanism in Globular Clusters}, (astro-ph/0211492)


\bibitem{LISA}
\url{ http://lisa.jpl.nasa.gov}

\bibitem{B02}
M.\ J.\ Benacquista, Class. Quantum Grav. {\bf 19}, 1297 (2002)

\bibitem{TB76}
K.\ S.\ Thorne and V.\ B.\ Braginsky, 
Astrophys.\ J.\, {\bf  204}, L1, (1976).

\bibitem{S_MOG}
S.\ Sigurdsson, Workshop summary for {\it Massive black holes
coalescence focus session}, Penn State, 2002, (gr-qc/0303027). 


\bibitem{BLS03}
O.\ Blaes, M.\ H.\  Lee and  A.\ Socrates, 
Astrophys.\ J.\, {\bf  578}, 775, (2002).

\bibitem{SIMT}
H.\ Sudou, S.\ Iguchi, Y.\ Murata and Y.\ Taniguchi, 
Science, {\bf 300}, 1263, (2003).

\bibitem{PM63} P. C. Mathews and J. Mathews, Phys.\
Rev.\ B {\bf 136}, 435 (1963).


\bibitem{W87} H. Wahlquist, Gen.\ Rel.\ Grav.\
{ \bf 19}, 1101, (1987).

\bibitem{LW90} C. W. Lincoln and C. M. Will, Phys.\
Rev.\ D {\bf 42},
1123 (1990).

\bibitem{GS89}
R.\ Wagoner and C.\ M.\ Will, 
Astrophys.\ J.\, {\bf  210}, 764, (1976);
erratum {\bf 215}, 984,(1977);
L. Blanchet and G. Sch\"afer,
Mon. Not. R. Astron.
Soc. {\bf 239}, 845 (1989);
W. Junker  and G. Sch\"afer, Mon. Not. R. Astron.
Soc. {\bf 254}, 146 (1992);
 L. Blanchet and G. Sch\"afer, Class. Quantum Grav.
{\bf 10}, 2699 (1993);
R. Rieth and G. Sch\"afer, Class. Quantum Grav.
{\bf 14}, 2357 (1997);


\bibitem{MBM95} C. Moreno-Garrido, J. Buitrago and E. Mediavilla,
Mon. Not. R. Astron.
Soc. {\bf 274}, 115 (1995).


\bibitem{MBM94} C. Moreno-Garrido, J. Buitrago and E. Mediavilla,
Mon. Not. R. Astron.
Soc. {\bf 266}, 16 (1994).


\bibitem{MP99} K. Martel and E. Poisson,
Phys.Rev. {\bf D60}, 124008 (1999).

\bibitem{PSLR}
V. Pierro, I. M. Pinto, A. D.  Spallicci, E. Laserra and F. Recano,
Mon. Not. R. Astron.
Soc. {\bf 325}, 358 (2001).

\bibitem{GIKPP} A. V. Gusev, V. B. Ignatiev, A. G.  Kuranov, K. A.  Postnov, 
and M. E. Prokhorov,
Astron.\ Lett.\, {\bf 28}, 143, (2002).

\bibitem{S01} N. Seto, Phys.\ Rev.\ Lett.\ {\bf 87}, 
251101 (2001).


\bibitem{BDIWW} L. Blanchet, T. Damour, B.R. Iyer,
C. M. Will and A. G. Wiseman, Phys.\ Rev.\ Lett.\ {\bf 74}, 3515
(1995);
L. Blanchet, T. Damour, and B. R. Iyer,  Phys.\
Rev.\ D
{\bf 51}, 5360 (1995).

\bibitem{BIWW96}
L. Blanchet, B.R. Iyer,
C. M. Will and A. G. Wiseman,
Class.\ Quantum Grav.\ {\bf 13},
 575 (1996).

\bibitem{DIJS_N}
T.\ Damour, B.\ R.\ Iyer, P.\  Jaranowski and  B.\ S.\ Sathyaprakash,
Phys.\ Rev.\ D {\bf 67}, 064028, (2003).


\bibitem{WW96} C. M. Will and A. G. Wiseman, Phys.\ Rev.\ D
{\bf 54}, 4813 (1996).

\bibitem{GI97} A. Gopakumar and B. R. Iyer,
Phys.\ Rev.\ D {\bf 56}, 7708 (1997).


\bibitem{DS88}T. Damour and G. Sch\"afer, Nuovo Cimento B
{\bf 101}, 127 (1988).

\bibitem{SW93}G.. Sch\"afer and N. Wex, Phys. Lett. {\bf 174 A},
196,
(1993); erratum {\bf 177}, 461.

\bibitem{GI02}
A.\ Gopakumar and  B.\ R.\ Iyer,
Phys.\ Rev.\ D {\bf 65}, 84011 (2002).

\bibitem{GI03er}
The 2PN corrections to the rate of decay of orbital elements 
of the generalized quasi-Keplerian representation,
computed in \cite{GI97} and the 2PN contributions 
to gravitational wave polarizations found in \cite{GI02}
are erroneous.
These errors were due to an incorrect implementation of 
the contact transformations between
the harmonic and ADM coordinates. As these coordinates are identical to
the 1PN order, errors surfaced at 2PN order  
terms in the ADM coordinates.
The errata, correcting  these errors, will be published soon.
In  Appendix \ref{AppA}, we present the {\em correct} 
transformations for
${\bf x}$ and ${\bf v}$ in these coordinates.

\bibitem{TD82}
T. Damour, in \emph{Gravitational Radiation}, N. Deruelle and T.
Piran (eds.), North-Holand Company (1983).


\bibitem{DD85} T. Damour and N. Deruelle, Ann. Inst. Henri Poincare
Phys. Theor. {\bf 43}, 107 (1985).

\bibitem{DT92} T. Damour and J. Taylor, Phys. Rev D {\bf  45 },  1840 (1992).


\bibitem{TD83}
T.\ Damour, 
Phys.\ Rev.\ Lett. \ {\bf 51}, 1019 (1983).

\bibitem{TD_F85}
T.\ Damour in {\em  Proceedings of Journ\'ees Relativistes} 1983,
edited by S.\ Benenti, M. Ferraris and M. Francaviglia
( Pitagora Editrice, Bologna, 1985), pp 89-110.

\bibitem{AI03} K. G.  Arun, L. Blanchet, B. R. Iyer and M. S. S. Qusailah,
{\it In Preparation}, (2004).

\bibitem{GS83}
G.\ Sch\"afer, Lettere Al  Nuovo Cimento, {\bf 36}, 105 (1983),
{\it ibid}, Ann. Phys. (N.Y.) \textbf{161}, 81 (1985);
T. Damour and G. Sch\"afer, Nuovo Cimento B
{\bf 101}, 127 (1988).



\bibitem{DJS00A}
P.\ Jaranowski and G.\ Sch\"afer,
Phys.\ Rev.\ D {\bf 57}, 7274 (1998); {\it ibid}, {\bf 60}, 124003 (1999);
T.\ Damour, P.\ Jaranowski, and G.\ Sch\"afer,
Phys.\ Rev.\ D 
{\bf 62}, 021501(R) (2000) [Erratum: {\bf 63}, 029903(E) (2001)];
{\it ibid} {\bf 63}, 044021, (2001)
[Erratum,  {\bf 66}, 029901(E), (2002)];
T.\ Damour, P.\ Jaranowski, and G.\ Sch\"afer,
Phys.\ Lett.\ B {\bf 513}, 147 (2001).

\bibitem{BF01}
L.\ Blanchet and G.\ Faye,
Phys.\ Lett.\ A {\bf 271}, 58 (2000);
Phys.\ Rev.\  D {\bf 63}, 062005 (2000);
J.\ Math.\ Phys.\ {\bf 41}, 7675 (2000); {\it ibid} {\bf 42}, 4391 (2001).

\bibitem{DJS00B}
T. Damour, P. Jaranowski and G. Sch\"afer,
Phys. Rev. D {\bf 62}, 044024, (2000).


\bibitem{ABF02}
V.\ C.\ de Andrade, L.\ Blanchet, and G.\ Faye, 
Class.\ Quantum Grav.\ {\bf 18}, 753,  (2001);

\bibitem {BI02} L. Blanchet and B. R. Iyer, 
Class.\ Quantum Grav.\ {\bf 20}, 755,  (2003).


\bibitem{DS85} T.\ Damour and G.\ Sch\"afer,
Gen.\ Rel.\ Grav.\ { \bf 17}, 879, (1985).



\bibitem{BD99} A. Buonanno and  T. Damour,
Phys.Rev. D {\bf 59},  084006 (1999);
{\it ibid} Phys.Rev. D {\bf 62},  064015 (2000).

\bibitem{DJS00}
T. Damour, P. Jaranowski and G. Sch\"afer,
 Phys.Rev. D {\bf 62},  084011 (2000).

\bibitem{GK02} K. Glampedakis and D. Kennefick,
Phys.Rev. D {\bf 66}, 044002 (2002).

\bibitem{E_def}
We point out a slight mismatch in the definition of $E$ used here and
in \cite{DS88,SW93}. We observe that 
$E$ employed in \cite{DS88,SW93}, say $E_{\rm DS}$, gives
the orbital (non-relativistic) energy per unit reduced mass, which in terms
of ${\cal E}$ reads $E_{\rm DS} \equiv \frac{{\cal{E}}-m c^2}{\mu }$.
In this paper, $E \equiv \frac{ {\cal E} -m c^2}{\mu\,c^2 }
= \frac{ E_{\rm DS}}{c^2} $.
Similarly, the dimensionless angular momentum 
variable  $j \equiv c\,{\cal{J}}/( \mu\, G\,m) $
is related to $h$, used in \cite{DS88,SW93},
by $j=h c$.


\bibitem{GS03} 
R.-M. Memmesheimer, A. Gopakumar,
S. Menzel, 
 and  G. Sch\"afer, {\it In Preparation}, (2004). 


\bibitem{GS85}
G. Sch\"afer, Ann. Phys. (N.Y.) \textbf{161}, 81 (1985);
B.\ R.\ Iyer and C.\ M.\ Will, Phys.\ Rev.\ D {\bf 52}, 6882 (1995);
C. K\"onigsd\"orffer, G. Faye and G. Sch\"afer
 Phys.\ Rev.\ D {\bf 68}, 044004 (2003).

\bibitem{BD88}  L. Blanchet and T. Damour, Phys. Rev. D
{\bf 37}, 1410 (1988); {\it ibid} {\bf 46}, 4304 (1992).

\bibitem{BS93}
L. Blanchet and G. Sch\"afer, Class. Quantum Grav.
{\bf 10}, 2699 (1993);

\bibitem{WW27}
W.H. Whittaker and G.N. Watson, 
{\it Modern Analysis}, CUP, Cambridge (1927). 

\bibitem{Def_Inst}
Following \cite{BDIWW}, we term contributions to the GW amplitude
and its associated quantities
that depend only on the state of the binary at the retarded
instant as its `instantaneous' part, whereas
the contributions, which are {\it a priori} sensitive to 
the entire `history' of the binary's dynamics are termed 
as the `tail' contributions.


\bibitem{RS97} 
R. Rieth and G. Sch\"afer, Class. Quantum Grav.
{\bf 14}, 2357 (1997).

\bibitem{W95}
N.\ Wex, 
Class.\ Quantum\ Grav.\ {\bf 12}, 983 (1995).

\bibitem{KG04}
C. K\"onigsd\"orffer and A. Gopakumar, {\it In Preparation}, (2004).

\end{thebibliography}
\end{document}